\documentclass[prb,aps,preprint]{revtex4}
\usepackage[latin1]{inputenc}
\usepackage[T1]{fontenc}
\usepackage{graphicx}
\usepackage{amsmath}
\usepackage{color}

\begin{document}

\bigskip

\title{Quantum master equations
for the superconductor--dot Entangler}
\author{Olivier Sauret, Denis Feinberg}
\affiliation{Laboratoire d'Etudes des Propri\'et\'es Electroniques des Solides, Centre National de la Recherche Scientifique, BP 166,38042 Grenoble Cedex 9, France}
\author{Thierry Martin}
\affiliation{Centre de Physique Théorique et  Université de la Méditerranée, Case 907, 13288 Marseille, France}

\begin{abstract}
The operation of a source of entangled electron spins, based on a superconductor and two quantum 
dots in parallel\cite{loss}, is described in detail with the help of quantum master equations. These are derived 
including the main parasitic processes in a fully consistent and non-perturbative way, starting from a microscopic
Hamiltonian. The average current is calculated, including the contribution of entangled and non-entangled
pairs. The constraints on the operation of the device are illustrated by a calculation of the various charge
state probabilities.
\end{abstract}

\maketitle

\section{Introduction}

Entanglement is a basic resource in quantum computation and quantum communication \cite{q_information}.
Recently, various experiments for quantum information processing schemes have been successfully 
implemented with photons as Bell inequality violation \cite{aspect} or teleportation\cite{bennett,bouw}. 

Any system with a two-level quantum degree of freedom is a possible candidate to carry a quantum bit. One of such 
is the electron and its spin. In principle, individual electrons can be manipulated in a quantum circuit and have 
the advantage of promising high-level integration in electronic devices \cite{loss_divincenzo}. Notice that the electron 
flow can be in principle much larger than the photon flow in equivalent optical devices where attenuation is necessary
to produce individual photons. Moreover, photons essentially do not interact except 
during their generation process, whereas Coulomb correlations between electrons in a quantum circuit open the 
possibility for new operations between quantum bits \cite{oliver,TP}. 

Non-locality in quantum mechanics can be 
probed by letting two quantum degrees of freedom interact, and subsequently separating these two systems. 
Here, electronic entanglement can be created using a superconductor\cite{loss,lesovik_martin_blatter}, where two 
electrons forming a Cooper pair are in a singlet state. The superconductor is coupled to two arms, each of 
them collecting one electron from each Cooper pair. The emission of one electron in each lead from the same Cooper 
pair corresponds to the so-called Crossed Andreev process \cite{torres,deutscher_feinberg,falci}, which can be
understood as a non-local Andreev reflection: the emission of one of the electrons can be seen as the absorption of
a hole with opposite spin and opposite momentum.
The two electrons forming the singlet are then spatially separated. It is then necessary to avoid the ``ordinary''
Andreev reflection where the two electrons go into the same lead. This selection can be enforced, either with the 
help of spin filters, leading to energy entanglement\cite{lesovik_martin_blatter}. Or alternatively, one can use 
energy filters, leading to spin entanglement\cite{loss,lesovik_martin_blatter}. Quantum dots with Coulomb blockade,
inserted in each branch, can efficiently select the crossed Andreev process.
As another possibility, the superconductor can be replaced by a normal 
quantum dot\cite{oliver,saraga_loss}. In this paper, the studied device consisting of a superconductor 
connected to two quantum dots in parallel will be called the Entangler (see Fig.\ref{Entangler}). 
Branching currents in the right and left leads were calculated for this Entangler in Ref. \onlinecite{loss} using a 
T-matrix approach. Entanglement can be probed by sending the electrons from a splitted pair into a 
beamsplitter\cite{buttiker} and by measuring noise correlations\cite{burkard}.

In the present paper a microscopic derivation of quantum master equations\cite{blum} for the Entangler is presented. 
It provides a simple, intuitive approach to probe entanglement and to monitor the effect of parasitic 
processes. Compared to a T-matrix derivation \cite{loss}, this approach has the advantage of describing the whole 
charge dynamics in a non-perturbative way (this statement will be qualified below). This allows to derive not only 
the average current but also the higher moments
of the current distribution. Another point is that quantum master equations can be applied  to any arbitrary
quantum system containing superconducting elements, or to another kind of Entangler.

Over the past years a great interest has been devoted to the description of the transport properties
through devices containing coupled nanostructures, where quantum interference has a strong influence.
A rather accessible method, generalizing the classical master equations\cite{sequential}, has been developed
in Ref.\onlinecite{gurvitz_prager} where Bloch-type quantum rate equations have been derived using 
the Schr\"odinger equation. When the system is an isolated quantum dot in the Coulomb blockade,
only the diagonal elements of the density matrix (the occupation probabilities) enter the rate 
equations. On the other hand, when the transfer of electrons through a quantum device goes through a
superposition of states in the different parts of this device, non-diagonal matrix elements will
appear in the equations of motion. The master equations then take into account coherent
processes and are a generalization of the Bloch equations \cite{cohen_dupont}.

The microscopic derivation of these equations provides a good understanding of the correspondence between
quantum and classical descriptions of transport in mesoscopic systems. The crucial point is the decoupling
between the time scales which specify, first, the dynamics inside the reservoirs and, secondly, the inverse 
rates for coupling the quantum states and the leads. This decoupling procedure is justified as long as the 
time scales characterizing transfer within the quantum system and injection (emission) from (to) the reservoirs 
are both large compared to the time scale for fluctuations within the reservoirs. This is equivalent to a 
markoffian hypothesis \cite{cohen_dupont}.

Until now, the generation of quantum master equations has been limited to
the case of sequential tunneling within quantum dots coupled to normal
reservoirs, using a microscopic Hubbard-type Hamiltonian \cite{gurvitz_prager,gurvitz}. In particular,
electron transfer which is mediated by high lying virtual states is not accounted for.
Consider now the case of a superconductor coupled to quantum dots:
single-electron tunneling does not conserve energy and is forbidden as the electron
transfer is accompanied by the emission of a Bogolubov quasiparticle.
However two-electron events such as Andreev processes (transfer of a pair of electrons out of the superconductor)
and superconducting cotunneling (S--cotunneling) processes\cite{cotunneling} (transfer of an electron from one
dot to another via the superconductor) connect low energy states, and thus enter the lowest-order contribution to the
tunneling current from the superconductor. One simplification would be to assume that the two-electron tunneling 
processes occur simultaneously, and are described by a pair Hamiltonian: rate equations have been written recently 
in this manner for the transport processes in a teleportation cell which employs an array of normal
and superconducting quantum dots \cite{TP}. However, in presence of transport channels mixing different processes,
it is safer to derive quantum master equations starting directly from the microscopic Hamiltonian. This is achieved 
in the present work, taking into account the main parasitic processes. The sequence of relevant steps will
clearly require virtual states which contribute to Andreev and cotunneling events.
After having established the equations including coherent quantum mechanical effects and Coulomb blockade,
we will determine their range of validity and show the relevance of the lifetime of virtual states.
The derivation of quantum master equation is non-perturbative with regard to transitions within the Entangler, while
the coupling to the leads is treated within the Fermi golden rule as in the orthodox theory of Coulomb blockade\cite{sequential}.

The paper is organized as follows. In Sec.\ref{model}, we present the system and its energy scales, together with
the Crossed Andreev process -- the main process -- and the important parasitic processes that can occur
during its evolution. This allows to write the many-excitation wavefunction which is the  starting point of
each microscopic derivation. This derivation is first described in Sec.\ref{ideal} for the Crossed
Andreev process, without any parasitic process. Parasitic processes are presented next, and compared 
in Sec.\ref{parasit} before giving the complete description of the system by quantum master equations in Sec.\ref{total} 
and appendix \ref{system}. Sec.\ref{discussion}
provides the physical discussion of the operation of the device as a function of its parameters.

%------------------------------------------- Model ----------------------------------------------------

\section{The Entangler device and its parameters}
\label{model}

\subsection{The model}

\begin{figure}[!hbt]

 \centering{  
 \includegraphics[width=7cm]{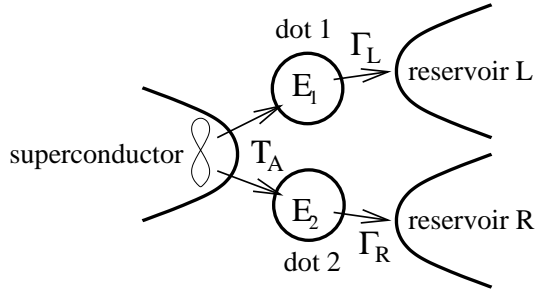} 

 \caption{\footnotesize{The Entangler setup: a superconductor injects electrons in quantum dots $D_1$ and $D_2$,
whose energies in state $|1\rangle$ (i.e. one excess electron) are respectively $E_1$ and $E_2$.
Electrons in the dots can subsequently tunnel into the normal reservoirs $L,R$.}}
 \label{Entangler}
 }
 \end{figure}

Let us first provide a qualitative description of the Entangler. The setup involves a superconductor ($S$) 
coupled by tunneling barriers to two quantum dots ($D_1$ and $D_2$) which are themselves coupled to normal 
leads, $L$ and $R$ (see Fig.\ref{Entangler}). Only one level is retained in each dot, assuming the level 
separation in each dot to be large enough\cite{loss}. The energy levels of the dots can be tuned by external 
gate voltages. The microscopic Hamiltonian of the entire system is the following:
\begin{equation} \label{hamiltonian}
\mathcal{H}=\mathcal{H}_0+\mathcal{H}_{tunnel}
\end{equation}
where
\begin{equation} \label{hamiltonian_cin}
\mathcal{H}_{0}=\sum_{k,\sigma}E_k\gamma_{k\sigma}^{\dagger}\gamma_{k\sigma}+E_1d_{1\sigma}^{\dagger}d_{1\sigma}+E_2d_{2\sigma}^{\dagger}d_{2\sigma}+U_1 n_{1\sigma}n_{1-\!\sigma}+U_2 n_{2\sigma}n_{2-\!\sigma}+\sum_{l\sigma}E_la_{l\sigma}^{\dagger}a_{l\sigma}+\sum_{r\sigma}E_ra_{r\sigma}^{\dagger}a_{r\sigma}
\end{equation}
where $\gamma_{k\sigma}$, $d_{i\sigma}$, $a_{l\sigma}$,  $a_{r\sigma}$ are destruction operators for Bogolubov 
quasiparticles, dot electrons and reservoir electrons. $n_{i\sigma}=d^{\dagger}_{i\sigma}d_{i\sigma}$ is the 
occupation number in the dots, which enters the Hubbard repulsion term with coupling constants $U_1$ and $U_2$.
A possible inter-dot repulsion is omitted here for sake of simplicity, but it could easily incorporated in the 
energies of various charge states of the two dots system.

The tunnel Hamiltonian which connects these elements by a one-electron transition reads:
\begin{equation} \label{hamiltonian_tunnel}
\mathcal{H}_{tunnel}=\sum_{k,\sigma}\Omega_{k1}d_{1\sigma}^{\dagger}c_{k\sigma}+\sum_{k,\sigma}\Omega_{-k2}d_{2-\!\sigma}^{\dagger}c_{-k-\!\sigma}+\sum_{l,\sigma}\Omega_{l}a_{l\sigma}^{\dagger}d_{1\sigma}+\sum_{r,\sigma}\Omega_{r}a_{r\sigma}^{\dagger}d_{2\!\sigma}+h.c.
\end{equation}

with a single electron tunneling amplitude $\Omega_1$ ($\Omega_2$) between $S$ and $D_1$ ($S$ and $D_2$),
and  $\Omega_l$ ($\Omega_r$) between $D_1$ and $L$ (between $D_2$ and $R$). $\sigma=\{\frac{1}{2},-\frac{1}{2}\}$ 
is the spin variable. Note that $\mathcal{H}_{tunnel}$ is written in the Fourier space. Point contacts are assumed 
between $S$ and dots $1$ and $2$ (in $\vec{r}_1$ and $\vec{r}_2$) thus the tunneling term is 
$\Omega_i d_{i\sigma}^{\dagger}c_{r\sigma}$\cite{loss}, 
which can be written in the Fourier space: $\sum_k\Omega_{i}e^{i\vec{k}.\vec{r}}d_{i\sigma}^{\dagger}c_{k\sigma} =
\sum_k\Omega_{ki}d_{i\sigma}^{\dagger}c_{k\sigma}$. The effective momentum dependence of the tunneling amplitude 
$\Omega_{ki}$ introduces a geometrical factor, which can strongly influence the transition amplitude for 
processes involving the two quantum dots.
During the injection process, Cooper pairs are initially separated into one electron in a dot and one 
quasiparticle in $S$. We introduce the Bogulubov transformation:
\begin{equation}
\left\{
\begin{array}{c}
c_{k\sigma}^{\dagger}=u_k\gamma_{k\sigma}^{\dagger}+\sigma
v_k\gamma_{-k-\!\sigma}S^{\dagger}\\
c_{k\sigma}=u_k^*\gamma_{k\sigma}+\sigma
v_k^*\gamma_{-k-\!\sigma}^{\dagger}S
\end{array}
\right.
\end{equation}
with
\begin{equation}
u_k=\frac{1}{\sqrt{2}}\left(1+\frac{\xi_k}{E_k}\right)^{\frac{1}{2}}
\end{equation}
\begin{equation}
v_k=\frac{1}{\sqrt{2}}\left(1-\frac{\xi_k}{E_k}\right)^{\frac{1}{2}}e^{i\phi_S}
\end{equation}
\begin{equation}
E_k=\sqrt{\xi_k^{\phantom{k}2}+\Delta^2}=\sqrt{\frac{\hbar^2k^2}{2m}-\mu_S+\Delta^2}
\end{equation}

Here $S$ stands for the annihilation of a Cooper pair \cite{tinkham}, and $\phi_S$ is the superconductor's phase. 
The two electrons from a Cooper pair become an entangled pair of electrons (only the singlet state is involved) when
going into different leads. Current flow is imposed by a voltage bias $\Delta\mu$ between the superconductor and the 
leads.
The basic mechanism for entanglement is based on a Crossed Andreev process between the superconductor and the
two quantum dots, forced by the Coulomb blockade in the dots. First, two entangled electrons are created in $D_1$ 
and $D_2$ via a virtual state which contains a quasiparticle in $S$ whose energy is larger than $\Delta_S$, the
superconducting gap. This process is coherent, and couples the superconducting chemical potential $\mu_S$ and the 
final energy of the pair in the dots $E_1+E_2$. This Andreev process probability is optimized at $E_1+E_2=\mu_S$, 
and behaves like a narrow two-particle Breit-Wigner resonance. Then the two electrons tunnel independently to each 
lead. This whole sequence of events forms the Crossed Andreev channel.

\subsection{Working conditions}
Next, the relevant parameters describing the device are discussed, following Ref.\onlinecite{loss}.
First, the charge states of the quantum system have to be well separated to avoid transitions due to thermal excitations.
But the thermal energy must be large enough in comparison to the transition probability to allow the markoffian
hypothesis. Therefore $\Gamma_{L,R}\ll k_B\Theta\ll E_i-\mu_{L,R}$. In order to conserve spin and thus the singlet 
state during the electron
transfer, spin-flip must be excluded. Thus each dot cannot carry a magnetic moment which could interfere with an
electron coming from $S$, i.e. it must carry an even number of electrons \cite{loss}.
Moreover, when an electron is deposited on a dot, another electron of this dot with opposite spin could in principle 
escape to the normal leads thus spoiling the entanglement. This spin-flip process is suppressed when the dot level 
spacing $\delta\epsilon$ is larger than the imposed bias $\Delta\mu$ and the temperature $k_B\Theta$.
Entanglement loss can also occur because of electron-hole excitations out of the Fermi sea of the leads
during the tunneling sequence. Such many-particle contributions are suppressed if the resonance width
$\Gamma_{L,R}=2\pi \rho_{L,R}(E_{1,2})|\Omega_{L,R}(E_{1,2})|^2$ is smaller than $E_{1,2}-\mu_{L,R}$.
This justifies the microscopic Hamiltonian of Eq.\ref{hamiltonian}.

Next, given this Hamiltonian, one needs to justify the derivation of the quantum master equation.
Single-electron tunneling from the superconductor to the leads via
the dots is avoided because it implies the creation of a quasiparticle in $S$. This process costs at 
least $\Delta_S$ which is assumed to be much larger than $\Delta\mu$ and $k_B\Theta$.

\subsection{Parasitic processes}
The main purpose of this device is to force the two electrons from a pair to propagate in the two different leads.
In a clean three-dimensional superconductor, this process is decreased by a geometrical factor
$\gamma_A=e^{-\frac{r}{\pi\xi_0}}\frac{\sin(k_Fr)}{k_Fr}$ ($\xi_0$ is the superconductor coherence length and 
$r=|\vec{r}_1-\vec{r}_2|$ is the distance of the two contacts between dots and $S$). 
The crossed Andreev amplitude is then $\gamma_A T$, with $T= (\pi /2)N(0)\Omega_1\Omega_2$.
Beside the decay on $\xi_0$, the algebraic factor can be improved by reducing the dimensionality\cite{recher_loss}
 or using a dirty superconductor\cite{feinberg}. Incidentally, the finite width of the contacts may introduce
diffraction corrections to the geometrical factor. Note that when taking into account the finite thickness of
the contacts, the geometrical factor can be modified\cite{prada}.

There are three main parasitic processes which could decrease the Entangler efficiency. Two of them create different 
channels of emission of two electrons coming from a Cooper pair, for which the two electrons can tunnel to the same lead
\cite{loss}. Although they involve higher energy intermediate states, those do not suffer from the geometrical factor of the 
Crossed Andreev channel. In addition, an elastic cotunneling -- this process will be called S--cotunneling in what follows -- 
connects every channel to other processes by transferring an electron between the two dots via $S$.

\begin{figure}[!hbt]

 \centering{  
 \includegraphics[width=\textwidth]{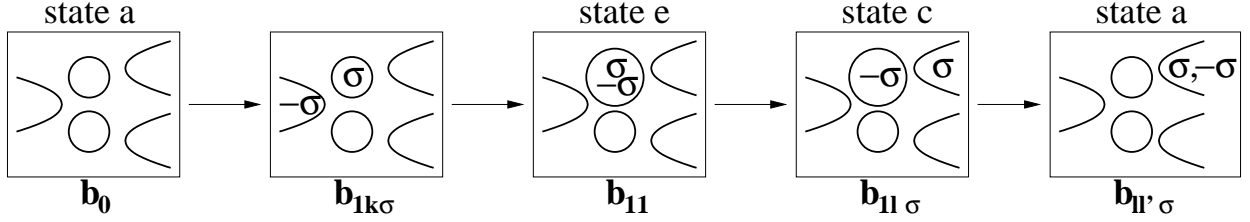} 

 \caption{\footnotesize{A current channel sending a pair of electrons to a same reservoir. Andreev
process towards one quantum dot can happen against strong Coulomb repulsion $U$.}}
 \label{blocage_coulomb}
 }
 \end{figure}

The two electrons of a Cooper pair can tunnel through the same dot by an Andreev process (cf. Fig.\ref{blocage_coulomb}).
Because of double occupancy, the pair would get an energy $U$ due to Coulomb repulsion. This is a coherent process 
between two energy levels with a large energy difference $U$. Because this energy cost is much larger than the 
Andreev process probability amplitude $T_i\sim N(0)\Omega_i^2$ involving a single lead, this process is strongly 
suppressed\cite{loss}. 
Alternatively, a pair could propagate to the same lead if the first electron injected on a given dot leaves it
before the second electron is deposited on either dots.  It goes to the corresponding lead while its twin electron
``has been staying in $S$'' as part of a quasiparticle (cf. Fig.\ref{branche}).
The latter can then choose toward which dot it will tunnel. It will prefer the same dot in order
to avoid paying the geometrical factor. This latter process costs $\Delta_S$ and thus can be
suppressed with $\Delta_S\gg \gamma_AT$. Let us notice that this process requires three transitions, including
one transition to a reservoir, thus it is not coherent.

\begin{figure}[!hbt]

 \centering{  
 \includegraphics[width=10.5cm]{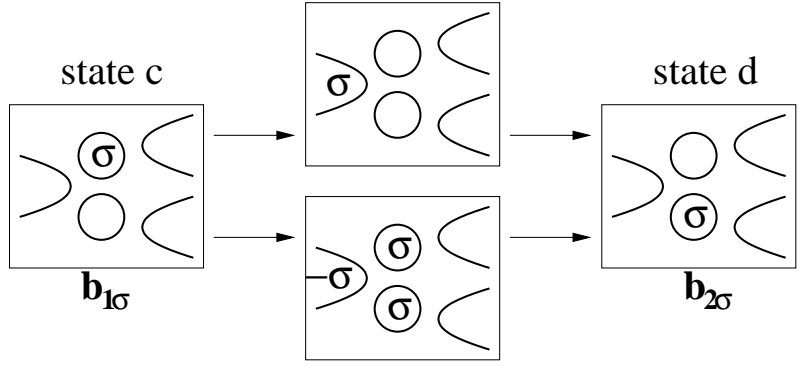} 

 \caption{\footnotesize{Cotunneling between the two dots. An electron from dot $1$ tunnels towards the dot $2$
via a virtual intermediate state containing a quasiparticle. Two contributions participate to the cotunneling
depending on when the initial electron is transferred.}}
 \label{cotunneling}
 }
 \end{figure}

By a S--cotunneling process via $S$, an electron can tunnel from $D_1$ ($D_2$) to $D_2$ ($D_1$)(cf. Fig. \ref{cotunneling}).
This is a coherent process between two discrete energy levels, $E_1$ and $E_2$ for a single
electron in the two dots or $U_1$ and $E_1+E_2$ for a doubly occupied dot ($U_2$ and $E_1+E_2$ for the
opposite configuration). Cotunneling is characterized by an amplitude $\gamma_CT$, with its own geometrical factor
$\gamma_C$. If the energy difference between the two coupled levels is much larger than the process amplitude $T_C$, 
this process will be weak.\\

To summarize, the working regime of the device is the following:
\begin{equation}
\Delta_S,U,|E_1-E_2|>\delta\epsilon>\Delta\mu,k_B\Theta>\Gamma_{L,R},T_A,T_C
\end{equation}
This working regime contains the justifications for the approximations made in the derivation of the 
master equation: the markoffian approximation and the relevant processes involving at most two successive 
virtual states with only one quasiparticle in $S$.

In what follows we also assume that $\Delta\mu> k_B \Theta$, in order to ensure the irreversibility of the pair 
production.

%-------------------------------------------entanglement process----------------------------------------------------

\section{Master equations for the crossed Andreev channel}
\label{ideal}

The transport channels which are described above can be characterized by the charge configuration
of the isolated quantum system for each step of the Entangler operation. The quantum
system is composed of the dots and the superconductor, but its dynamics can be directly probed by
integrating out excitations in the reservoirs and superconductor. Using the Schr\"odinger equation and
generalizing the procedure of ref.\onlinecite{gurvitz_prager}, it is shown here how to derive
quantum master equations which describe the evolution of the reduced density matrix of the system.
As a starting point, we consider the dynamics in the situation where only the Crossed Andreev process and
one-electron relaxation processes are effective -- the ideal regime. The wavefunction is thus chosen
to include only the charge states involved in this particular channel. A reduced Hilbert space containing 
the lowest energy states and the required virtual intermediate states is chosen
(containing a single quasiparticle in $S$).

 \begin{figure}[!hbt]

 \centering{  
 \includegraphics[width=\textwidth]{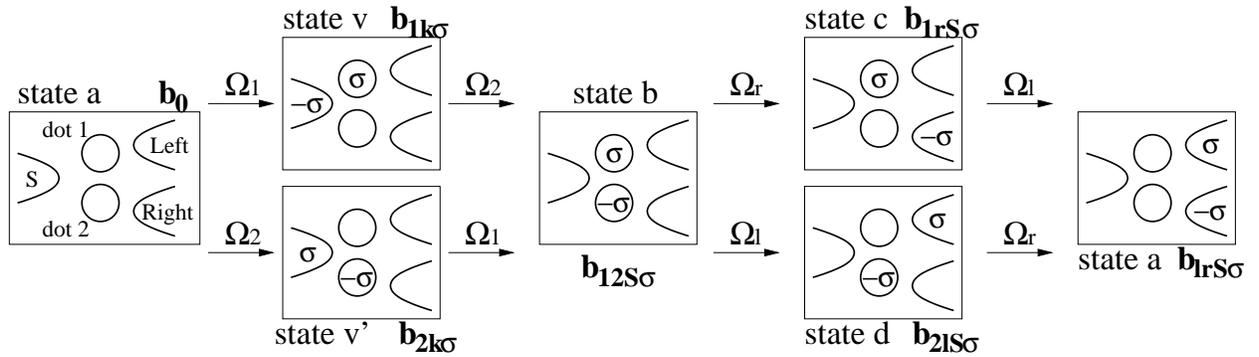} 

 \caption{\footnotesize{Sequence of states for the Crossed Andreev channel of the Entangler.
For instance, $b_{ik,\sigma}$ denotes the amplitude to have an electron in dot $i$
while a quasiparticle is created in the superconductor. First an electron is deposited in either dot, 
next the second electron tunnels and forms a singlet state in the pair of dots, next either electron is absorbed in
the reservoir, and finally the two dots are empty.}}
 \label{sequence}
 }
 \end{figure}

The many-excitation wavefunction for this problem is written as:
\begin{equation} \label{fctonde}
\left|\Psi(t)\right\rangle=\left[
\begin{array}{cl}
b_0(t)&+\sum_{k,\sigma}b_{1k,\sigma}(t)d_{1\sigma}^{\dagger}\gamma_{-k-\!\sigma}^{\dagger}S+\sum_{k,\sigma}b_{2k,\sigma}(t)d_{2-\!\sigma}^{\dagger}\gamma_{k\sigma}^{\dagger}S\\
&+\sum_{\sigma}b_{12S,\sigma}(t)d_{1\sigma}^{\dagger}d_{2-\!\sigma}^{\dagger}S\\
&+\sum_{l,\sigma}b_{2lS,\sigma}(t)a_{l\sigma}^{\dagger}d_{2-\!\sigma}^{\dagger}S
+\sum_{r,\sigma}b_{1rS,\sigma}(t)d_{1\sigma}^{\dagger}a_{r-\!\sigma}^{\dagger}S\\
&+\sum_{lr,\sigma}b_{lrS,\sigma}(t)a_{l\sigma}^{\dagger}a_{r-\!\sigma}^{\dagger}S\\
\\
&+\sum_{lr,\sigma,k',\sigma'}b_{lrS\sigma,1k'\sigma'}(t)d_{1\sigma'}^{\dagger}\gamma_{-k'-\!\sigma'}^{\dagger}a_{l\sigma}^{\dagger}a_{r-\!\sigma}^{\dagger}SS'\\
&+\sum_{lr,\sigma,k',\sigma'}b_{lrS\sigma,2k'\sigma'}(t)d_{2-\!\sigma'}^{\dagger}\gamma_{k'\sigma'}^{\dagger}a_{l\sigma}^{\dagger}a_{r-\!\sigma}^{\dagger}SS'\\
&+\qquad \cdots\qquad +\qquad \cdots\qquad +\qquad \cdots
\end{array}
\right]|0\rangle
\end{equation}
where $b_{...}(t)$ are the time-dependent amplitudes for finding the system in the corresponding states
with the initial conditions $b_0(0)=1$ and all other $b(0)$ are zero. The indices indicate the electron 
occupation in the dots and reservoirs, as depicted in Fig. \ref{sequence}. 
The use of Schr\"odinger equation and the form of $|\Psi(t)\rangle$ call for some comments. In fact, as said above, 
the temperature is not zero thus one should in principle rely on a density matrix description from the beginning. 
Yet, under the condition $\Gamma_{L,R}<k_B\Theta<\Delta\mu$, one can simply use the Schr\"odinger equation in a 
reduced subspace of states\cite{gurvitz_prager}. Those states for instance do not include electron-hole excitations in 
the same reservoir: these are supposed to relax on a very short time, due to inelastic processes occurring in $L$ 
and $R$. On the contrary, all possible charge and spin states on the dots, together with all excitations including 
holes in $L$ and electrons in $R$, are considered. Summing on these reservoir states eventually lead to the equations 
for the two reduced density matrix\cite{gurvitz_prager}.

After substituting  Eq.(\ref{fctonde}) into the Schr\"odinger equation $i|\dot{\psi}(t)\rangle=\mathcal{H}|\psi(t)\rangle$,
an infinite set of coupled linear differential equations is obtained for $b(t)$ by projecting
$i\langle\psi_i|\dot{\psi}(t)\rangle=\langle\psi_i|\mathcal{H}|\psi(t)\rangle$ for each state $|\psi_i\rangle$.
$|\psi_i\rangle$ characterizes the quantum state of the total system including the environment. Applying the Laplace transform
\begin{equation} \label{1a}
\tilde{b}(E)=\int_0^{\infty}e^{i(E+i\eta)t}b(t)\;dt
\end{equation}
and taking into account the initial conditions, an infinite set of
algebraic equations is obtained for the amplitudes $\tilde{b}(E)$ (see Fig.\ref{sequence}):
\begin{subequations} \label{1}
	\begin{gather}
(E+i\eta)\tilde{b}_{0}-\sum_{k\sigma}\sigma
v_k\Omega_{k1}^*\tilde{b}_{1k\sigma}+\sum_{k\sigma}\sigma
v_k\Omega_{-k2}^*\tilde{b}_{2k\sigma}=i \label{1a} \\
(E+i\eta-E_1-E_k)\tilde{b}_{1k\sigma}=\sigma
v_k^*\Omega_{k1}\tilde{b}_{0}+u_k\Omega_{-k2}^*\tilde{b}_{12S\sigma} \label{1b} \\
(E+i\eta-E_2-E_k)\tilde{b}_{2k\sigma}=-\sigma
v_k^*\Omega_{-k2}\tilde{b}_{0}-u_k\Omega_{k1}^*\tilde{b}_{12S\sigma} \label{1c} \\
(E+i\eta-E_1-E_2)\tilde{b}_{12S\sigma}=\sum_{k}u_k^*\Omega_{-k2}\tilde{b}_{1k\sigma}-\sum_{k}u_k^*\Omega_{k1}\tilde{b}_{2k\sigma}+\sum_{r}\Omega_r\tilde{b}_{1rS\sigma}+\sum_{l}\Omega_l\tilde{b}_{2lS\sigma} \label{1d} \\
(E+i\eta-E_1-E_r)\tilde{b}_{1rS\sigma}=\Omega_r\tilde{b}_{12S\sigma}+\sum_l\Omega_l\tilde{b}_{lrS\sigma} \label{1e} \\
(E+i\eta-E_2-E_l)\tilde{b}_{2lS\sigma}=\Omega_l\tilde{b}_{12S\sigma}+\sum_r\Omega_r\tilde{b}_{lrS\sigma} \label{1f} \\
(E+i\eta-E_l-E_r)\tilde{b}_{lrS\sigma}-\sum_{k'\sigma'}\sigma'
v_k\Omega_{k1}^*\tilde{b}_{lrS\sigma,1k'\sigma'}+\sum_{k'\sigma'}\sigma'
v_k\Omega_{-k2}^*\tilde{b}_{lrS\sigma,2k'\sigma'}=\Omega_l\tilde{b}_{1r\sigma}+\Omega_r\tilde{b}_{2l\sigma} \label{1g} 
	\end{gather}
\end{subequations}
$$\ldots$$
Each term corresponds to the transition between two successive states. Each transition leads to the
creation or annihilation of a quasiparticle either in $S$ or in a reservoir. There is an fundamental difference between the
two types of transitions. The first one involve an excited state whose lifetime is so small ($\tau_{qp}\sim 1/\Delta_S\ll 1/T$)
that coherence is kept until the quasiparticle is destroyed. On the other hand, in the reservoirs, quasiparticles
instantaneously decay ($\tau_{relax}\sim 1/E_F\ll 1/\Gamma$) so coherence is lost (Markoff process).
To simplify the system of equations,  the expression for $\tilde{b}$ is substituted in terms of the type 
$\sum\Omega\tilde{b}$ from equations containing sums. Every sum over the continuum states ($k$, $l$, $r$) 
is replaced by integrals (see Appendix \ref{integrale}). Crossed terms (like $\sum_l\tilde{b}_l\Omega_l\Omega_r/(E-E_l)$) 
vanish\cite{gurvitz_prager}, and the following set of equations is obtained:
\begin{subequations} \label{2}
\begin{gather}
(E+i\eta-2c(T_1+T_2))\tilde{b}_{0}=2\gamma_ATe^{i\phi_S}(\tilde{b}_{12S,\sigma}-\tilde{b}_{12S,-\sigma}) \label{2a} \\
(E+i\eta-E_1-E_2-c'(T_1+T_2)+i\frac{\Gamma_L}{2}+i\frac{\Gamma_R}{2})\tilde{b}_{12S,\sigma}=2\sigma\gamma_ATe^{-i\phi_S}\tilde{b}_{0}\label{2b} \\
(E+i\eta-E_1-E_r+i\frac{\Gamma_L}{2})\tilde{b}_{1rS,\sigma}=\Omega_r\tilde{b}_{12S,\sigma}\label{2c} \\
(E+i\eta-E_2-E_l+i\frac{\Gamma_R}{2})\tilde{b}_{2lS,\sigma}=\Omega_l\tilde{b}_{12S,\sigma}\label{2d} \\
(E+i\eta-E_l-E_r-2c(T_1+T_2))\tilde{b}_{lrS,\sigma}=2\gamma_ATe^{i\phi_S}
\left(\tilde{b}_{lrS\sigma,12S'\sigma'}-\tilde{b}_{lrS\sigma,12S'-\!\sigma'}\right)+\Omega_l\tilde{b}_{1rS,\sigma}+\Omega_r\tilde{b}_{2lS,\sigma}\label{2e}
\end{gather}
\end{subequations}
$$\ldots$$
with $T_i = \frac{\pi}{2}N(0) \Omega_i^2$ and $c$, $c'$ are numerical constants (see Appendix \ref{integrale}), 
involved in self-energy corrections. Here the coefficients for virtual states (states $|v\rangle$ and 
$|v'\rangle$ in Fig.\ref{sequence}) have disappeared from the equations. This is the consequence of the 
succession of quasiparticle creation and annihilation transitions forced by the assumption that two 
quasiparticles cannot coexist in $S$.

The singlet/triplet basis is now chosen. For instance, in the global wave function,
$\sum_{\sigma}\tilde{b}_{12S,\sigma}d^{\dagger}_{1\sigma}d^{\dagger}_{2-\!\sigma}$ is replaced by
$\tilde{b}_{12S}^{singlet}(d_{1\sigma}^{\dagger}d_{2-\!\sigma}^{\dagger}-d_{1-\!\sigma}^{\dagger}d_{2\sigma}^{\dagger})/\sqrt{2}+\tilde{b}_{12S}^{triplet}(d_{1\sigma}^{\dagger}d_{2-\!\sigma}^{\dagger}+d_{1-\!\sigma}^{\dagger}d_{2\sigma}^{\dagger})/\sqrt{2}$.
From Eq.(\ref{2b}) one can say that coefficients $\tilde{b}_{12S,\sigma}$
and $\tilde{b}_{12S,-\sigma}$ for a given spin are opposite.
This is the same for $\tilde{b}_{1rS,\sigma}$ and $\tilde{b}_{1rS,-\sigma}$, $\tilde{b}_{2lS,\sigma}$ and
$\tilde{b}_{2lS,-\sigma}$. 
The tunnel Hamiltonian conserves spin, therefore there is no coupling towards triplet spin states.
Thus $\tilde{b}_{ijS}^{singlet}=\sqrt{2}b_{ijS,\sigma}=-\sqrt{2}b_{ijS,-\!\sigma}$ and $\tilde{b}_{ijS}^{triplet}=0$.

The density matrix elements of the set-up are now introduced.
The Fock space of the quantum dots consists of four possible charge states:
$|a\rangle$ - levels $E_1$ and $E_2$ are empty, $|b\rangle$ - levels $E_1$
and $E_2$ are occupied,
$|c\rangle$ - level $E_1$ is occupied, $|d\rangle$ - level $E_2$ is
occupied. Reservoirs states are identified by $n$,
the number of pairs of electrons out from $S$ to the reservoirs. 
To obtain the reduced density matrix, elements are summed over $n$:
\begin{equation}
\sigma_{\alpha\beta}=\sum_{n=0}^\infty
\sigma_{\alpha\beta}^{(n)}
\end{equation}

In every state, electrons are paired in a singlet state.
The matrix elements are defined as:
\begin{eqnarray*}
\sigma_{aa}&=&{\displaystyle\left|\tilde{b}_0(t)\right|^2+\sum_{l,r}\left|\tilde{b}_{lrS}^{singlet}\right|^2+\sum_{l<l',r<r'}\left|\tilde{b}_{lrS,l'r'S'}^{singlet}\right|^2+\cdots} \\
\sigma_{bb}&=&{\displaystyle\left|\tilde{b}_{12S}^{singlet}\right|^2
+\sum_{l',r'}\left|\tilde{b}_{l'r'S',12S}^{singlet}\right|^2+\cdots} \\
\sigma_{cc}&=&{\displaystyle\sum_r\left|\tilde{b}_{1rS}^{singlet}\right|^2+\sum_{l',r'<r}\left|\tilde{b}_{l'r'S',1rS}^{singlet}\right|^2+\cdots} \\
\sigma_{dd}&=&{\displaystyle\sum_l\left|\tilde{b}_{2lS}^{singlet}\right|^2+\sum_{l'<r,r'}\left|\tilde{b}_{l'r'S',2lS}^{singlet}\right|^2+\cdots}  \\
\sigma_{ab}&=&{\displaystyle \tilde{b}_0\tilde{b}_{12S}^{singlet*}+\sum_{l,r}\tilde{b}_{lrS}^{singlet}\tilde{b}_{lrS,12S'}^{singlet*}+\cdots} \\
\sigma_{ba}&=&\sigma_{ab}^*
\end{eqnarray*}

The matrix density elements are directly related to the coefficients $\tilde{b}(E)$ by a Laplace transform:
\begin{equation}
\sigma_{\alpha\beta}^{(n)}=\sum_{l\ldots,r\ldots}\int\frac{dE\,dE'}{4\pi^2}\tilde{b}_{l\ldots,r\ldots}(E)\tilde{b}_{l\ldots,r\ldots}^*(E')
\end{equation}
where $\alpha/\beta$ specify the charging states associated with the
amplitudes ($b$'s).
The equations for $n=0$ can be obtained straightforwardly. For instance,
to get $\sigma_{aa}^{(0)}$, Eq.(\ref{2a}) is multiplied by
$\tilde{b}_0^*(E')$ and the conjugate equation written for $E'$
is subtracted.
\begin{subequations}\label{tralala}
\begin{gather}
\dot{\sigma}_{aa}^{(0)}=2\sqrt{2}i\gamma_AT\left(e^{-i\phi_S}\sigma_{ab}^{(0)}-e^{i\phi_S}\sigma_{ba}^{(0)}\right) \label{tralala1} \\
\dot{\sigma}_{bb}^{(0)}=-\left(\Gamma_L+\Gamma_R\right)\sigma_{bb}^{(0)}-2\sqrt{2}i\gamma_AT\left(e^{-i\phi_S}\sigma_{ab}^{(0)}-e^{i\phi_S}\sigma_{ba}^{(0)}\right) \\
\dot{\sigma}_{cc}^{(0)}=-\Gamma_L\sigma_{cc}^{(0)}+\Gamma_R\sigma_{bb}^{(0)} \\
\dot{\sigma}_{dd}^{(0)}=-\Gamma_R\sigma_{dd}^{(0)}+\Gamma_L\sigma_{bb}^{(0)} \\
\dot{\sigma}_{aa}^{(1)}=2\sqrt{2}i\gamma_AT\left(e^{-i\phi_S}\sigma_{ab}^{(1)}-e^{i\phi_S}\sigma_{ba}^{(1)}\right)+\Gamma_L\sigma_{dd}^{(0)}+\Gamma_R\sigma_{cc}^{(0)}\label{tralala2} 
\end{gather}
\end{subequations}
$$\ldots$$
Note that the diagonal matrix elements (the ``populations'') are coupled with the off-diagonal
density-matrix elements (``coherences''), which is symptomatic of a coherent, reversible transition. 

To obtain the equations for the coherence one subtracts Eq.(\ref{2a}) for $E$
multiplied by $\tilde{b}_{12s,singlet}^*(E')$ and Eq.(\ref{2b}) for $E'$ multiplied by $\tilde{b}_0^*(E)$:
\begin{equation}
\dot{\sigma}_{ab}^{(0)}=-\frac{1}{2}\left(\Gamma_L+\Gamma_R\right)\sigma_{ab}^{(0)}+i\left(E_1+E_2+K(T_1+T_2)\right)\sigma_{ab}^{(0)}+2\sqrt{2}i\gamma_ATe^{i\phi_S}\left(\sigma_{aa}^{(0)}-\sigma_{bb}^{(0)}\right)
\end{equation}
where $K=c'-2c$. 

These equations describe the sequential evolution of the system and involve consequently only processes between real states. 
Coherent processes (not involving reservoirs) couple non-diagonal elements to diagonal elements while relaxation processes 
couple only diagonal elements. From the set of equations \ref{tralala}, one can see that these processes do not 
interfere because of the loss of phase coherence introduced by the markoffian approximation, i.e. the sum over reservoir 
states. A density matrix element for one particular state is then only coupled to the elements for adjacent states in the 
sequence. Thus the processes can be added easily, which will be crucial when considering the full operation including all 
channels.

Here, because only one current channel is implied in the ideal operation, we
can easily verify that the equations are the same for each $n$. Therefore the sum over $n$ is obvious and
one obtains the master equations for the evolution of the density matrix describing the system:
\begin{subequations}\label{system}
\begin{gather}
\dot{\sigma}_{aa}=2\sqrt{2}i\gamma_AT\left(e^{-i\phi_S}\sigma_{ab}-e^{i\phi_S}\sigma_{ba}\right)+\Gamma_L\sigma_{cc}+\Gamma_R\sigma_{dd} \label{system1} \\
\dot{\sigma}_{bb}=-2\sqrt{2}i\gamma_AT\left(e^{-i\phi_S}\sigma_{ab}-e^{i\phi_S}\sigma_{ba}\right)-\left(\Gamma_L+\Gamma_R\right)\sigma_{bb} \label{system2} \\
\dot{\sigma}_{cc}=-\Gamma_L\sigma_{cc}+\Gamma_R\sigma_{bb} \label{system3} \\
\dot{\sigma}_{dd}=-\Gamma_R\sigma_{dd}+\Gamma_L\sigma_{bb} \label{system4} \\
\dot{\sigma}_{ab}=-\frac{1}{2}\left(\Gamma_L+\Gamma_R\right)\sigma_{ab}+i\left(E'_1+E'_2\right)\sigma_{ab}+2\sqrt{2}i\gamma_ATe^{i\phi_S}\left(\sigma_{aa}-\sigma_{bb}\right) \label{system5}
\end{gather}
\end{subequations}
with $E'_i=E_i+KT_i$.

This is the main result of this section. First, let us remark that the transition rates $\Gamma_{L,R}$ appear only from
the dots to the reservoirs, and not in the opposite direction. This is consistent with the assumption that $k_B\Theta$ 
is small compared to the transition energies between dots and reservoirs. This limitation of Gurvitz's method is not a 
problem here since the Entangler actually needs to be strongly biased to avoid decoherence effects.
The second term of Eq.(\ref{system5}) expresses that two discrete energy
levels are coupled by a coherent process  involving two transitions. Note that the probability 
of transmission between these two states is maximum in the resonant case, e.g. $\varepsilon=E'_1+E'_2$ is zero.\\

The ideal operation of the system involves only one channel for transferring a Cooper pair to the reservoirs:
the two electrons tunnel towards different leads. Actually, using the normalization condition for the populations
$\sigma_{aa}+\sigma_{bb}+\sigma_{cc}+\sigma_{dd}=1$, equations (\ref{system}) are easily solved for
the stationary current, $I=I(t\rightarrow\infty)$ ($\dot{\sigma}_{\alpha\beta}=0$):
\begin{equation} \label{entanglement_current}
I_L^{ent}/e=\Gamma_L\sigma_{bb}+\Gamma_L\sigma_{cc}=\frac{\Gamma_L\Gamma_R}{\Gamma_L+\Gamma_R}\;\;\frac{8\gamma_A^2T^2}{8\gamma_A^2T^2+\frac{\Gamma_L\Gamma_R}{4}+\varepsilon^2\frac{\Gamma_L\Gamma_R}{(\Gamma_L+\Gamma_R)^2}}
\end{equation}
\begin{equation}
I_R^{ent}=e\Gamma_R\sigma_{bb}+e\Gamma_R\sigma_{dd}=I_L^{ent}
\end{equation}
This current is made of entangled singlet pairs. This result was obtained earlier in Ref.\onlinecite{loss} in the
limit $\gamma_A T\ll\Gamma$ and $\Gamma_L=\Gamma_R$. Here the presence of the term $8\gamma_A^2T^2$ in the denomination 
comes from a complete (non-perturbative) treatment of both Andreev and decay processes.

The equality of the currents in the two branches of the device is a direct consequence 
of the Crossed Andreev process. Every electron pair crosses and goes out of the system - each electron on its own side -
before the next pair is injected in the dots. Those ``cycles'' never overlap in this ideal working regime.

In the case $\Gamma\gg \gamma_A T$, one obtains:
\begin{equation}
I_L^{ent}/e=\frac{8\gamma_A^2T^2(\Gamma_L+\Gamma_R)}{\varepsilon^2+(\Gamma_L+\Gamma_R)^2/4}
\end{equation}
while in the case $\Gamma\ll \gamma_A T$,
\begin{equation}
I_L^{ent}/e=\frac{\Gamma_L\Gamma_R}{\Gamma_L+\Gamma_R}
\end{equation}
like a single quantum dot between two leads\cite{sequential}.
In the latter situation, the dots are almost always occupied, so that 
the resistance is dominated by the rate associated with
the two barriers - in parallel - between dots and leads.

%--------------------------------------------------Parasitic processes-------------------------------------------------------

\section{Parasitic channels} \label{parasit}

The ideal working regime is affected by parasitic processes:
Andreev tunneling via a single dot, one-by-one tunneling or S--cotunneling.
The two first ones have been separately computed by the T-matrix in Ref.\onlinecite{loss}.
Their effect is to create different channels of pair current which decrease the efficiency of
entanglement. As said before, the terms for each process can be added in the
equations and combined before including them together in a whole system
of quantum master equations collecting every possible processes (see part \ref{total}).
To start with, the different processes will be separately considered.

%--------------------------------------------------Coulomb blockade--------------------------------------------------

\subsection{Direct Andreev effect process against Coulomb blockade}

Let us imagine that a Cooper pair tunnels to the same quantum dot by an
Andreev process, while generating a doubly occupied state.

Because of Coulomb repulsion, an energy $U$ (Eq.(\ref{hamiltonian_cin})) is required
for having two electrons in a same quantum dot. If $U$ is ``large
enough'', such a process will have a low probability. With conventional dot technology,
the interaction energy $U\sim1K$ in the quantum dots can be controlled so that it
is smaller than the superconducting gap $\Delta_S>2K$. Therefore the doubly occupied energy level has no coupling
to the continuum of $S$ quasiparticles, which would effectively introduce a broadening.

Similarly to the case of the ideal working regime (Sec.\ref{ideal}), the set of differential equations 
associated with this Direct Andreev channel are established for the reduced density matrix 
elements. Here, only one branch - say $L$ - is considered for simplicity. The Fock space of the quantum 
dots consists here of three possible charge states:
$|a\rangle$ - both dots are empty, $|e\rangle$ dot $1$ is doubly occupied,
$|c\rangle$ - dot $1$ is singly occupied. The wavefunction takes the
following form:
\begin{equation} \label{fctonde_Coul_bloc}
\left|\Psi(t)\right\rangle=\left[
\begin{array}{cl}
b_0(t)&+\sum_{k,\sigma}b_{1k,\sigma}(t)d_{1\sigma}^{\dagger}\gamma_{-k-\!\sigma}^{\dagger}S\\
&+\sum_{\sigma}b_{11}(t)d_{1\sigma}^{\dagger}d_{1-\!\sigma}^{\dagger}S\\
&+\sum_{l,\sigma}b_{1lS,\sigma}(t)a_{l\sigma}^{\dagger}d_{1-\!\sigma}^{\dagger}S\\
&+\sum_{l<l',\sigma}b_{ll'S,\sigma}(t)a_{l\sigma}^{\dagger}a_{l'-\!\sigma}^{\dagger}S\\
&+\sum_{l<l',\sigma,k',\sigma'}b_{ll'S\sigma,1k'\sigma'}(t)d_{1\sigma'}^{\dagger}\gamma_{-k'-\!\sigma'}^{\dagger}a_{l\sigma}^{\dagger}a_{l'-\!\sigma}^{\dagger}SS'\\
&+\qquad \cdots\qquad +\qquad \cdots\qquad +\qquad \cdots
\end{array}
\right]|0\rangle
\end{equation}
From the Schr\"odinger equation, and performing steps similar to Sec.\ref{ideal}, the set of equations for the density
matrix elements is:
\begin{subequations} \label{syst_coul_bloc}
\begin{gather}
\dot{\sigma}_{aa}=2iT_1\left(e^{-i\phi_S}\sigma_{ae}-e^{i\phi_S}\sigma_{ea}\right)+\Gamma_L\sigma_{cc} \label{syst_coul_bloc1}\\
\dot{\sigma}_{ee}=-2iT_1\left(e^{-i\phi_S}\sigma_{ae}-e^{i\phi_S}\sigma_{ea}\right)-2\Gamma'_L\sigma_{ee} \label{syst_coul_bloc2} \\
\dot{\sigma}_{cc}=-\Gamma_L\sigma_{cc}+2\Gamma'_L\sigma_{ee} \label{syst_coul_bloc3}\\
\dot{\sigma}_{ae}=i\left[U_{1}+K'T_1\right]\sigma_{ae}+2iT_1e^{i\phi_S}\left(\sigma_{aa}-\sigma_{ee}\right)+\Gamma'_L\sigma_{ae}\label{syst_coul_bloc4}
\end{gather}
\end{subequations}
with $K'$ a numerical constant, and $\Gamma'_L=2\pi \rho_L(U_{1}+E_1)|\Omega_L(U_{1}+E_1)|^2$ the level
broadening introduced by coupling of the two-electrons level with lead $L$. These equations are similar to 
equations (\ref{system}). Nevertheless, the sequence passes through a high energy-level 
($U$) via an Andreev process which implies an oscillation between two discrete energy levels, $\mu_S$ and
$U'_1=U_{1}+K'T_1$. On the contrary, in the ideal regime, this energy difference can be as small as desired.

%--------------------------------------------------Same branch--------------------------------------------------

\subsection{One-by-one electron tunneling to the reservoir} \label{branche}

This channel is another way to send a pair into one single lead. Before the second electron of
a broken Cooper pair can tunnel to a dot, the first one already leaves the dot to the corresponding 
lead. The second electron will tunnel through the same dot as its twin electron with a much higher probability
(Fig.\ref{branche}) than through the other dot, because of the geometrical factor. The latter process will be 
simply neglected.

 \begin{figure}[!hbt]

 \centering{  
 \includegraphics[width=\textwidth]{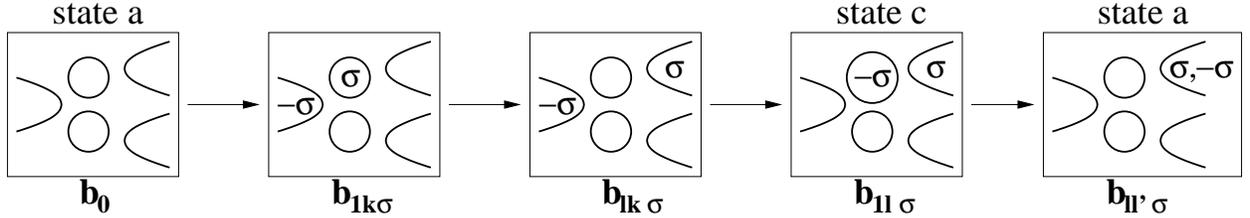} 

 \caption{\footnotesize{Sequence corresponding to the tunneling of a
singlet pair through one branch of the device. States $|a\rangle$ and $|c\rangle$
are coupled through two successive virtual states.}}
 \label{branche}
 }
 \end{figure}

There are only two processes involved in this channel. The first one, between states $|a\rangle$ 
and $|c\rangle$, requires two consecutive virtual states, both containing a quasiparticle in $S$.
Because of the coupling with a continuum of states in the lead, phase coherence
is lost thus off-diagonal matrix elements -- or coherences -- are not coupled to populations.
Therefore this channel is peculiar in the sense that it is incoherent even though it
involves transitions with $S$. The equations describing the evolution of the density matrix are 
obtained as before. The Schr\"odinger equation gives:

\begin{subequations}
\begin{gather}
\left(E+i\eta\right)\tilde{b}_0=i+\sum_{k,\sigma}\sigma v_k\Omega_{k1}^*\tilde{b}_{1k\sigma} \\
\left(E+i\eta-E_1-E_k\right)\tilde{b}_{1k\sigma}=\sigma
v_k^* \Omega_{k1}\tilde{b}_0+\sum_l \Omega_{l}^*\tilde{b}_{lk\sigma}\\
\left(E+i\eta-E_l-E_k\right)\tilde{b}_{lk\sigma}=\Omega_l\tilde{b}_{1k\sigma}+u_k
\Omega_{k1}^*\tilde{b}_{1l\sigma} \\
\left(E+i\eta-E_{1}-E_l\right)\tilde{b}_{1l\sigma}=\sum_k
u_k^*\Omega_{k1}\tilde{b}_{lk\sigma}+\sum_{l'}\sigma
\Omega_{l'}\tilde{b}_{ll'\sigma}\\
\left(E+i\eta-E_l-E_{l'}\right)\tilde{b}_{ll'\sigma}=
\Omega_{l'}\tilde{b}_{1l\sigma}-\Omega_{l}\tilde{b}_{1l'-\!\sigma}+\sum_{k,\sigma'}\sigma'
v_k \Omega_{k1}^*\tilde{b}_{ll'\sigma,1k\sigma'}
\end{gather}
\end{subequations}

Let us eliminate $\tilde{b}_{lk\sigma}$. To simplify, the notation
$\Delta_{ij}=E+i\eta-E_i-E_j$ is introduced.
\begin{equation}
\left(E+i\eta-2c\Omega_1^2\right)\tilde{b}_0=i+\sum_{k,l,\sigma}\frac{\sigma u_k v_k |\Omega_{k1}|^2\Omega_l}{\Delta_{lk}(\Delta_{1k}+i\Gamma_L/2)}\tilde{b}_{1l\sigma}
\end{equation}
\begin{equation}
\left(\Delta_{1l}-c''\Omega_1^2+i\frac{\Gamma_L}{2}\right)\tilde{b}_{1l\sigma}=\sum_k\frac{\sigma u_k^*v_k^*|\Omega_{k1}|^2\Omega_l}{\Delta_{lk}(\Delta_{1k}+i\frac{\Gamma_L(\Delta)}{2})}\tilde{b}_0+\sum_{kl'}\frac{|u_k|^2|\Omega_{k1}|^2\Omega_{l'}}{\Delta_{lk}\Delta_{l'k}(\Delta_{1k}+i\frac{\Gamma_L(\Delta)}{2})}\tilde{b}_{1l'\sigma}
\end{equation}
Finally, using integrals calculated in Appendix \ref{Abranche}, eq. \ref{one-by-one}, the
following set of equations is obtained:
\begin{eqnarray}
&\displaystyle{\left(E+i\eta-2c\Omega_1^2+i\left(\frac{2T_1}{\pi \Delta_S}\right)^2\Gamma_L\right)\tilde{b}_0=i}&\\
&\displaystyle{\left(\Delta_{1l}-c''\Omega_1^2+i\frac{\Gamma_L}{2}\right)\tilde{b}_{1l\sigma}=\sigma \frac{2T_1}{\pi \Delta_S}e^{-i\phi_S} \Omega_l \tilde{b}_0+\sigma \frac{3T_1}{2\pi \Delta_S} \Omega_1^2\sum_{l'}\Omega_{l'}\tilde{b}_{1l'\sigma}}& \\
&\displaystyle{\left(\Delta_{ll'}-2c\Omega_1^2+i\left(\frac{2T_1}{\pi \Delta_S}\right)^2\Gamma_L\right)\tilde{b}_{ll'\sigma}=\Omega_{l'}\tilde{b}_{1l\sigma}-\Omega_{l}\tilde{b}_{1l'\sigma}}&
\end{eqnarray}
Virtual states have disappeared from the equations. The remaining term in $T_1\Omega_l$ corresponds to the
three-step process coupling $|a\rangle$ to $|c\rangle$. Introducing the elements of the density matrix one gets :
\begin{subequations}
\begin{gather}
\dot{\sigma}_{aa}=-K''\frac{T_1^2}{\Delta_S^2}\Gamma_L\sigma_{aa}+\Gamma_L\sigma_{cc} \\
\dot{\sigma}_{cc}=-\Gamma_L\sigma_{cc}+K''\frac{T_1^2}{\Delta_S^2}\Gamma_L\sigma_{aa} \\
\end{gather}
\end{subequations}
where $K''=4/\pi^2$. This process behaves as the transport through a single dot where the
first barrier between the left lead and the dot is a three-step process via two virtual
states and the second barrier is a classic tunnel barrier.

%--------------------------------------------------Cotunneling--------------------------------------------------

\subsection{S--cotunneling between the two quantum dots}

Another process involves intermediate virtual states of the quantum device which are common to the other processes:
cotunneling\cite{cotunneling} between the two quantum dots via $S$. This process involves oscillations between two
position states and connects all of the channels studied until now.

It can occur in different situations: between states containing only one or two electrons in the
two dots. Like for the Crossed Andreev process, the transmission probability depends on the energy difference 
between the two coupled states. The equation of evolution for the density matrix describing oscillations
between two states -  $|c\rangle$ (electron on dot $1$)  and  $|d\rangle$ (electron on dot $2$) are established:
\begin{subequations}
\begin{gather}
\dot{\sigma}_{cc}=i\gamma_CT\left(\sigma_{cd}-\sigma_{dc}\right)\\
\dot{\sigma}_{dd}=i\gamma_CT\left(\sigma_{dc}-\sigma_{cd}\right)\\
\dot{\sigma}_{cd}=i\Delta E\sigma_{cd}+i\gamma_CT\left(\sigma_{cc}-\sigma_{dd}\right)\\
\end{gather}
\end{subequations}
where $\Delta E=E_2-E_1$, $\gamma_C=e^{-\frac{r}{\pi\xi_0}}\frac{\cos(k_Fr)}{k_Fr}$ is
the geometrical factor corresponding to this cotunneling process\cite{falci}. Note that when the distance
$r$ which separates the two tunneling locations is zero, $\gamma_C$ diverges.
This is expected because this process has no meaning for the same tunneling location:
this local process brings back the system in the same state, it only participates to the
renormalization of the energy level of the state by coupling with the continuum of quasiparticles in $S$. 
Note that the transition amplitude $T$ is the same as for Andreev process.

%------------------------------------------------tout d'un coup---------------------------------------------------

\section{Entangler in the presence of parasitic processes}
\label{total}

One of the advantage of Bloch-type equations is to be able
to study all processes together and non-perturbatively. In the previous sections, a specific
system of dynamical equations was obtained separately for different channels of pair current.
In particular, such channels are repeated cycle after cycle, which allows
to systematically group the contributions with different reservoir variables
(by recurrence over the number of pairs transmitted to the leads).
 
\begin{figure}[!hbt]

 \centering{  
 \includegraphics[width=\textwidth]{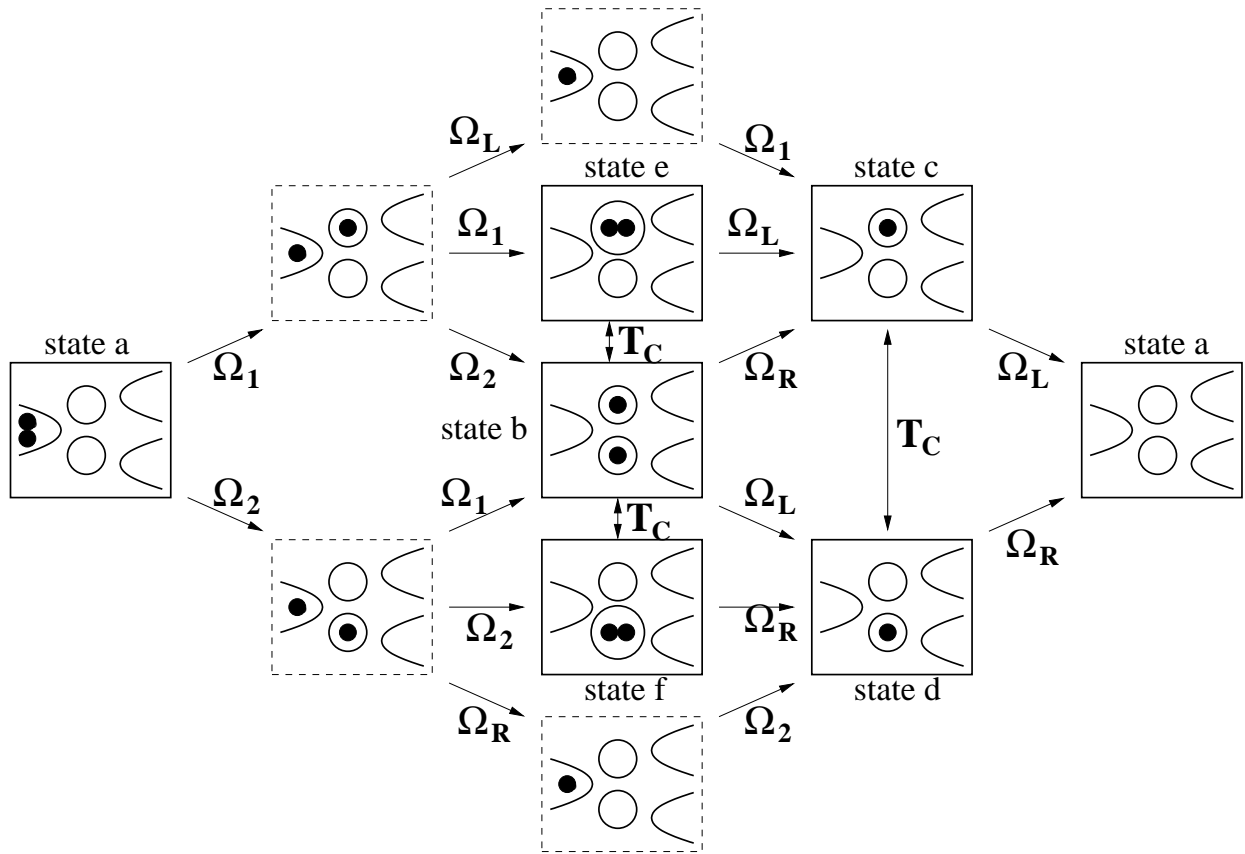} 

 \caption{\footnotesize{General operation including the three Andreev channels and S--cotunneling. States with three or 
four electron states are omitted for clarity. Real states are fully squared while virtual states are dashed squared. To 
make it simpler, spin is not represented. The $\Omega$'s correspond to transitions between two quantum states 
$|S\rangle\otimes|dots\rangle\otimes|l,r\rangle$, while $T_C$ indicates the resonant cotunneling process. Certain 
mixing process, like Direct Andreev effect between states $|c\rangle$ and state with one electron in dot $1$ and two 
electrons in dot $2$, are not presented for lack of space. However such processes are included in the quantum master 
equations.}}
 \label{general_operation of the Entangler}
 }
 \end{figure}

In reality, each channel (induced by Crossed-Andreev, Direct Andreev, S-cotunneling) 
mixes into one another, so one needs to gather all transitions in a single set of equations
for the density matrix. Because of this mixing, it is no more possible to establish a set
of equations cycle after cycle. 

A starting point for deriving generalized quantum master equations is thus to label the amplitude associated with each 
process by variables which count how many entangled pair have passed through reservoir $R$ or $L$ or both (while 
being split). Note that such variables do not appear in the quantum master equation of each channel because they 
have been summed over. It is straightforward, but tedious, to write a full Schr\"odinger equation for the most 
general operation, combining all states, and to derive the density matrix equations. The basic assumption is that 
not more than one quasiparticle is excited in the superconductor during the processes.

As was said in part \ref{ideal}, all the processes can be gathered without appealing to the full derivation 
of the Schr\"odinger equation, by adding terms corresponding to each process. We set the equations for a given
state of reservoirs were $n_L$ ($n_R$) singlet pairs of electrons have tunneled to the reservoir $L$ ($R$) and
$n_0$ pairs whose electrons have tunneled to different leads. States are thus defined by the charge of
quantum dots, one electron already in the reservoir while its twin electron is still in the quantum system, and
$n_L$, $n_R$ and $n_0$: $|\psi\rangle=|dot1,dot2\rangle\otimes|ll'rr'\rangle\otimes|n_L,n_R,n_0\rangle$. To get 
equations for only the charge states of the dots, they are summed over $l$, $l'$, $r$, $r'$ and the recurrence is 
made over $n_L$, $n_R$, $n_0$. The obtained set does not depend on the number of channels and leads. Thus Gurvitz's
method for generating quantum master equations\cite{gurvitz_prager} can be generalized to the multi-terminal 
case with many current channels. The full system is given in Appendix \ref{system}. One can notice that the
parasitic processes may generate triplet pairs in the leads $L,R$.

%------------------------------------------------Discussion-------------------------------------------------------------

\section{Discussion}
\label{discussion}

The set of quantum master equations will now be used to describe more quantitatively the transport properties.
To assess the constraints on parameters, each channel will first be studied, before using the complete set to
obtain a numerical evaluation of the operation in a realistic regime.

By solving quantum master equations one can find the average current for each uncoupled channel. This will be done 
for the symmetric case ($\Gamma_L=\Gamma_R$, $U_1=U_2$ and $T_1=T_2$) and assuming that $\Gamma_{L,R}=\Gamma'_{L,R}=\Gamma$ 
when the coupling between quantum dots and lead depends weakly on the energy:
The Direct-Andreev current is computed in the stationary regime with Eqs.(\ref{syst_coul_bloc}):
\begin{equation}
I_L^{Andreev}/e=2\Gamma'_L\sigma_{bb}+\Gamma_L\sigma_{cc}=\Gamma\frac{16T^2}{16T^2+\Gamma^2+U'^2}
\end{equation}
With $U'\gg 2T,\Gamma$ we have, as in Ref.\onlinecite{loss}
\begin{equation} \label{current_Cb_approx}
I_L^{Andreev}\approx e\Gamma\frac{16T^2}{U'^2}
\end{equation}
while with $\gamma_A^2T^2\gg\varepsilon^2$ Eq.(\ref{entanglement_current}) can be written:
\begin{equation} \label{current_ent_approx}
I_L^{CAndreev}\approx e\Gamma\frac{8\gamma_A^2T^2}{8\gamma_A^2T^2+\Gamma^2/4}
\end{equation}
The current created by the one-by-one tunneling process is given by:
\begin{equation} \label{current_sb_approx}
I_L^{sb}=e\frac{K''T^2\Gamma}{\Delta_S^2+K''T^2}\approx
4e\Gamma\frac{T^2}{\pi^2\Delta_S^2}
\end{equation}

Without taking here into account elastic cotunneling, one can see here the relationship between parameters that
must be fulfilled to approach the ideal working of the Entangler: $U',\Delta_S\gg \max[T,\Gamma/\gamma_A,\varepsilon/\gamma_A]$. 
This can be understood with a dynamical study of each channel. Actually Andreev processes are coherent processes
which create an oscillation between the state where the Cooper pair is in $S$ and states where the pair of electrons
is in the dots. Thus it will be a competition between the period and the amplitude of oscillations and the probability
of tunneling from a dot to a reservoir. Let first consider the case where $\gamma_AT\gg\Gamma$. Then for resonant 
Crossed-Andreev process
\begin{equation}
\sigma_{bb}(t)=\frac{1}{2}\left(1-\frac{\Gamma^2}{2T^2}\right)\left(1-\cos(2\gamma_A Tt)\right)e^{-\Gamma t}
\label{oscillationCA_thermo}
\end{equation}
while for Direct-Andreev process
\begin{equation}
\sigma_{ee}(t)=\frac{2T^2}{U'^2+4T^2}\left(1-\cos(\sqrt{U'^2+4T^2}t)\right)e^{-\Gamma t}
\label{oscillationDA_thermo}
\end{equation}
Because $\Gamma$ is small we are here in the regime where the Crossed-Andreev channel is more probable 
than the Direct-Andreev one because many oscillations between coherent states can occur before a transition to a reservoir has 
happened. On the other hand ($\gamma_AT\ll\Gamma$) one gets:
\begin{equation}
\sigma_{bb}(t)=\frac{\gamma_A^2T^2}{\Gamma^2}\left(e^{-2\gamma_A^2T^2/\Gamma t}+e^{-2\Gamma t}-2e^{-\Gamma t}\right)
\label{oscillationCA_cinetic}
\end{equation}
\begin{equation}
\sigma_{ee}(t)=\frac{T^2}{U'^2+\Gamma^2}\left(e^{-\Gamma T^2/(U'^2+\Gamma^2)t}+e^{-2\Gamma t}-2\cos(Ut)e^{-\Gamma t}\right)
\label{oscillationDA_cinetic}
\end{equation}
As soon as the pair has tunneled to the dots, it goes to the reservoirs. And because the Direct-Andreev frequency is 
larger than the Crossed-Andreev one ($U\gg\gamma_AT$), there is a small time interval in which Direct-Andreev is 
favored even though the amplitude of oscillation (and thus tunneling between $|a\rangle$ and $|e\rangle$) is smaller: 
for a relaxation time $1/\Gamma$ of the order of half a period of oscillation for Direct Andreev effect ($\pi/U$), 
after a time $t\sim 1/\Gamma$, the population of state $|e\rangle$ can be much larger than population of state $|b\rangle$. 

The same kind of argument can be given to study the effect of S--cotunneling. As said before, for $U$ and 
$\Delta_S$ large enough, the only parasitic effect is elastic cotunneling. Using only this process and Crossed-Andreev
process in the master equation,
the efficiency of entanglement is calculated depending on $E=|E_1-E_2|$ which controls S--cotunneling probability. We want 
to know the proportion of electrons from a same pair tunneling to different reservoirs ($P_{entangled}$) or to the same 
reservoir ($P_{parasitic}$). Cycles of current do not overlap so the probability is the same for each cycle. To calculate 
them, we can use Bloch equations describing the evolution on only one cycle to get first $|c\rangle$ and $|d\rangle$ 
populations as a function of time. From state $|b\rangle$ the first electron tunnels for example towards the left reservoir. 
The chance for the second electron to tunnel towards the right (left) reservoir is $\Gamma_Rp_d(t)$ ($\Gamma_Lp_c(t)$) 
assuming that $p_d(0)=1$. Thus $P_{entangled}=\int_0^{\infty}\Gamma_Rp_d(t)dt$. For $\Gamma_L=\Gamma_R$:
\begin{equation}
P_{entangled}=\frac{\Gamma^2+E^2+2\gamma_c^2T^2}{\Gamma^2+E^2+4\gamma_c^2T^2}
\label{efficiency}
\end{equation}
From equation (\ref{efficiency}), we can see that the condition to neglect S--cotunneling, leading to $P_{entangled}\sim 1$, 
is $\gamma_cT\ll \max[E,\Gamma]$.

\begin{figure}[!hbt]
 \centering{  
 \includegraphics[width=10.5cm]{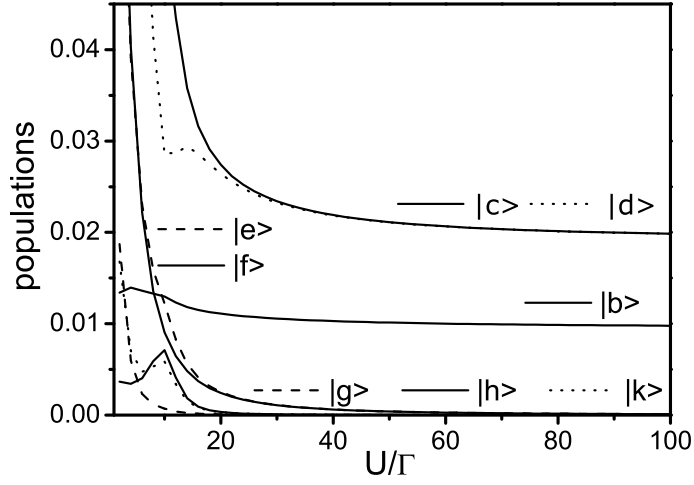} 
 \caption{\footnotesize{Charge states populations as a function of $U$ for $\Delta_S=9.5K$, $E_1=-E_2=0.5K$, 
$\Gamma_{L,R}=\Gamma'_{L,R}=T=0.1K$, $\gamma_A,\gamma_C\sim0.2$. States 
$|a\rangle$, $|b\rangle$, $|c\rangle$, $|d\rangle$, $|e\rangle$, $|f\rangle$ 
refer to Fig.\ref{general_operation of the Entangler}. State $|k\rangle$ refers to the triplet
state shared between dots, and states $|g\rangle$ and $|h\rangle$ refer to three electrons states 
(cf. Appendix \ref{system}). $|g\rangle$, $|h\rangle$ and $|k\rangle$, populations correspond to the three 
lowest curves. The population of states containing doubly occupied dots vanishes when $U$ increases. For 
low values of $U$ ($U\sim |E_1-E_2|$), the asymmetry is introduced by energy difference between states 
$|e\rangle$ (two electrons in dot $1$) and $|f\rangle$ (two electrons in dot $2$).}}
  \label{populationsU}}
 \end{figure}

\begin{figure}[!hbt]
 \centering{  
 \includegraphics[width=10.5cm]{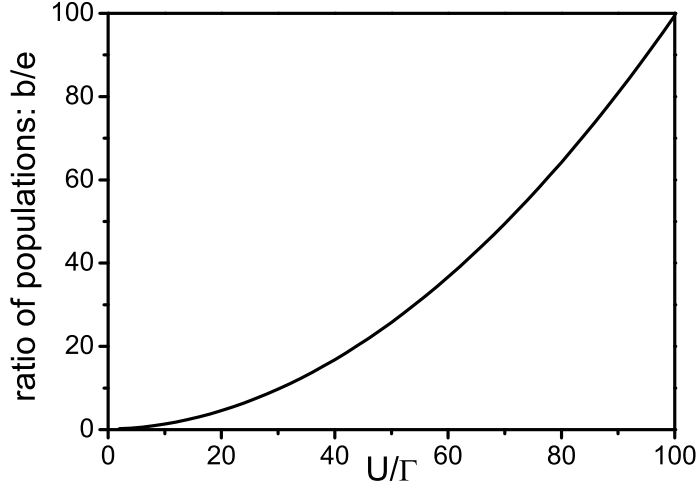} 
 \caption{\footnotesize{Ratio between populations of state $|b\rangle$ (singlet state shared between dots $1$ and $2$) 
and of state $|e\rangle$ population(two electrons in dot $1$) for $\Delta_S=9.5K$, $E_1=-E_2=0.5K$, 
$\Gamma_{L,R}=\Gamma'_{L,R}=T=0.1K$, $\gamma_A,\gamma_C\sim0.2$. It indicates the ratio between Direct-Andreev channel and 
Crossed-Andreev channel. The latter is strongly favored when $U$ increases ($p_b/p_e(U/\Gamma=90)=83.3$).}}
\label{ratio_cb_U} }
 \end{figure}

\begin{figure}[!hbt]

 \centering{  
 \includegraphics[width=10.5cm]{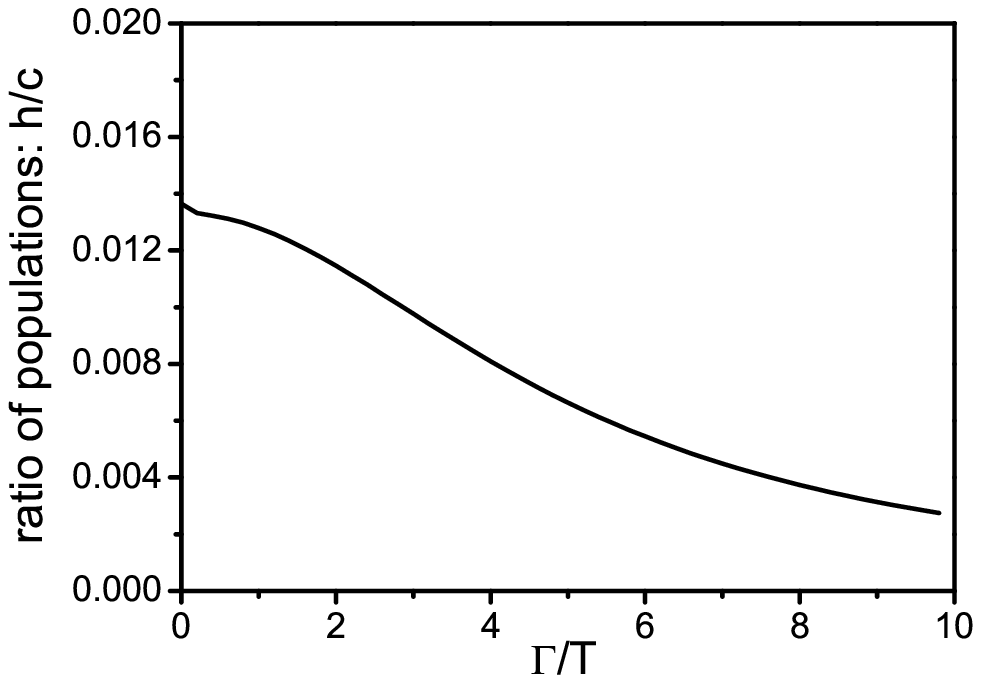} 

 \caption{\footnotesize{Ratio between populations of state $|h\rangle$ (one electron in dot $1$, two electrons 
in dot $2$) and of state $|c\rangle$ (one electron in dot $1$) for $\Delta_S=9.5K$, $E_1=-E_2=0.05K$, 
$\Gamma_{L,R}=\Gamma'_{L,R}$, $T=0.1K$, $U=1K$, $\gamma_A,\gamma_C\sim0.2$. Increasing $\Gamma$ compared to the
transition rate of Direct-Andreev and Crossed-Andreev processes allows to favor the decay of single charge states 
before another Cooper pair tunnels to the free quantum dot. For $U=1K=10T$, $p_h/p_e<1.5\%$.}}
 \label{ratio_he_G}}
 \end{figure}

A more general study using the complete set of equations (see Appendix \ref{system}) has to be performed. 
This set of equations can be solved in the stationary regime, but the general solution is typically cumbersome. 
For the sake of readability, it is presented here taking into account the parasitic processes only to first 
order. This fixes the different energy scales, previously discussed above, which define the working regime 
of the Entangler. Here the asymmetry $\Gamma_L\neq\Gamma_R$ is kept to show the role of S--cotunneling.

\begin{equation}
I_L=e\Gamma_L\sigma_{bb}+e\Gamma_L\sigma_{ee}+2e\Gamma'_L\sigma_{cc}
\end{equation}
\begin{equation} \label{courant}
\begin{array}{rl}
I_L=e\sigma_0&\left[\Gamma_L+\Gamma_R+4\Gamma^2\left(\frac{1}{\Gamma_L}-\frac{1}{\Gamma_R}\right)\frac{\gamma_C^2T^2}{\Delta E^2}\right.\\
&-2K''A\sigma_0\Gamma_L\frac{T^2}{\Delta_S^2}\left(1-\frac{1}{2\sigma_0}\right)\\
&-8A\sigma_0\Gamma_L\frac{T^2}{U^2}\left(2+\frac{\Gamma'_L}{\Gamma_L}\left(1-\frac{1}{\sigma_0}\right)+\frac{2\Gamma'_R}{\Gamma_R}\right)\\
&\left. -2\sigma_0\Gamma_L\frac{\gamma_C^2T^2}{U^2}\left(2+\frac{\Gamma'_L}{\Gamma_L}\left(1-\frac{1}{\sigma_0}\right)+\frac{2\Gamma'_R}{\Gamma_R}+\frac{2\Gamma\Gamma'_L}{\Gamma^2+8\gamma_A^2T_A^2+\varepsilon^2}\right)\right]
\end{array}
\end{equation}
where $A=\frac{8\gamma_A^2T_A^2}{\Gamma^2+8\gamma_A^2T_A^2+\varepsilon^2}$ and
$\sigma_0^{-1}=A+1+\Gamma_R/\Gamma_L+\Gamma_L/\Gamma_R$. From Eq.(\ref{courant}) we can exhibit which
parameters are controlling each contribution to the total current.

To complete this discussion, the set of equations is used to describe the average populations of each state
depending on some relevant parameters. $\Delta_S$ is taken to be the largest energy scale.
With niobium as superconductor, one takes $\Delta_S\sim9.25K$. For a two-dimensional quantum dot, small enough
($10$nm$^2$), one takes $|E_i|\sim0.5K$ and $U\sim9K$, with $T,\Gamma\sim0.1K$ \cite{kouwenhoven} and $\gamma_A,\gamma_C\sim0.2$.

On figure \ref{populationsU}, it can be seen that the population of states containing doubly occupied dots
vanishes when $U$ increases. It is important to notice that when $U\sim |E_i|$, the system is asymmetric 
and the channel with $U\sim |E_1-E_2|$ is favored because a Direct-Andreev process becomes resonant. 
At the working point ($U/T=90$) $p_e/p_b=0.012$. Two channels can be compared in calculating the ratio between 
two populations: on figure \ref{ratio_cb_U} the ratio between the population of state $|b\rangle$ and the 
one of state $|e\rangle$ indicates which of Crossed-Andreev and Direct-Andreev process is the most likely 
depending on $U$. Thus increasing $U$ increases the efficiency of entanglement. For small $U\sim\Gamma$,
the two channels become comparable because decays to reservoirs are much faster than Crossed-Andreev
oscillations.

A large $\Gamma$ will allow a fast transition between dots and reservoirs. That is why increasing $\Gamma/T$ 
will favor the most likely process which connects the superconductor to the dots\cite{loss}. 
On figure \ref{ratio_he_G} we can see that increasing $\Gamma$ favors the decay of a single charge state 
before another pair tunnels towards the free quantum dot. Actually, because 
Direct-Andreev oscillations are faster (frequency$\sim\sqrt{E^2+4T^2}$) than Crossed-Andreev oscillations 
(frequency$\sim\gamma_aT$), even if their probability is smaller, the decay towards reservoirs can happen
before one Crossed-Andreev process has been achieved. Thus increasing $\Gamma/T$ at fixed $U$, Direct-Andreev
process increase to the detriment of Crossed-Andreev one.

\section{Conclusion}

In this article, quantum master equations have been derived, starting from a microscopic Hamiltonian for the 
superconducting-dot Entangler. Using the Schr\"odinger equation technique developed in Ref.\onlinecite{gurvitz_prager},
the full equations describing the evolution of the reduced density matrix are obtained, retaining as virtual 
states only single particle excitations in the superconductor. Considering only one level by dot, all possible 
processes are taken into account in a fully consistent and non-perturbative way: Crossed-Andreev process,
responsible for entanglement, as well as Direct-Andreev and one-by-one tunneling processes, and cotunneling
through the superconductor. The latter connects all the other processes, yet the quantum master equations written
in Appendix \ref{system} take into account all processes in a coherent way. From them, the average current has 
been calculated.
The conditions on the Entangler parameters, needed for an optimal operation of the device, have been derived, 
and extend the result of Ref.\onlinecite{loss}.

The power of master equations is to give access, not only to the first moment, but to all moments of the current 
distribution\cite{bagrets}. In a forthcoming paper\cite{inpreparation}, shot noise correlations are computed in 
order to give a clear diagnosis of entanglement\cite{lesovik_martin_blatter,chtchelkatchev}. Another extention of Bloch equations 
is to include explicitly spin/charge relaxation or coupling to external degrees of freedom, in order to 
quantitatively study decoherence effects.

Such a derivation of quantum master equations, including higher order process, can obviously be generalized to a
wide class of quantum systems involving discrete charge states and coherent/incoherent transitions. It is
therefore a valuable tool for investigating nanostructures in view of controlling quantum information based 
on spin/charge degrees of freedom.
 
\acknowledgements
LEPES is under convention with Universit\'e Joseph Fourier. Funding from CNRS ``Action Concert\'ee Nanosciences''
is gratefully acknowledged.

%-------------------------------------------------------appendix-------------------------------------------------------------

\appendix
\section{Calculation of integrals} \label{integrale}

To obtain the evolution equation of the density matrix, it is necessary to compute some integrals
arising from the coupling between $S$ and the two dots.

\subsection{Crossed-Andreev effect}

The tunneling of the two electrons of a same Cooper pair to two different dots gives a contribution (see Eqs. \ref{1a}-\ref{1g}).
\begin{equation}
I_A=\sum_k\frac{u_kv_k\Omega_{k1}\Omega_{-k2}}{E-E_i-E_k}
\end{equation}
The two energy levels of the dots are assumed to be close to $\mu_S$.
The transitions amplitudes, $\Omega$, depend weakly on the
energy so they can be considered as constant with a phase factor
$e^{i\vec{k}.\vec{r}}$.  Neglecting $E-E_i\ll E_k\sim \Delta_S$
one obtains:
\begin{eqnarray}
I_A&=&{\displaystyle
-\Omega_{1}\Omega_{2}\frac{V}{(2\pi)^3}\int\,d^3\vec{k}\frac{\Delta}{2E_k^2}e^{i\vec{k}.\vec{r}}}\nonumber\\
&=&{\displaystyle
-\frac{\Omega_{1}\Omega_{2}}{2}\frac{V}{(2\pi)^3}\int_0^{2\pi}d\phi\int_0^{\pi}d\theta\int_{0}^{\infty}dk\,k^2\sin\theta e^{ikr\cos\theta}\frac{\Delta}{\Delta^2+(\frac{\hbar^2k^2}{2m}-\mu)^2}} \label{and3}
\end{eqnarray}
with $\mu=\frac{\hbar^2k_F^2}{2m}$ and $V$ the volume.\\
Because of parity the integral can be extended from $-\infty$ to $\infty$.
\begin{equation}
I_A=\frac{\pi\Omega_{1}\Omega_{2}}{2ir}\frac{V}{(2\pi)^3}\int_{-\infty}^{\infty}dk\,k\left(e^{ikr}-e^{-ikr}\right)\frac{\Delta}{\Delta^2+\left(\frac{\hbar^2k^2}{2m}-\frac{\hbar^2k_F^2}{2m}\right)}
\end{equation}
The four poles are:
$$
k=\pm k_F\sqrt[4]{1+\left(\frac{2m\Delta}{\hbar^2k_F^2}\right)^2}e^{\pm\frac{i}{2}\arctan\left(\frac{2m\Delta}{\hbar^2k_F^2}\right)}\equiv\pm ak_Fe^{\pm i\theta}
$$
$k_1=ak_Fe^{i\theta}$, $k_2=-ak_Fe^{-i\theta}$, $k_3=-ak_Fe^{i\theta}$,
$k_1=ak_Fe^{-i\theta}$.
the contour is the positive half-circle for $e^{ikr}$ and the
negative one for $e^{-ikr}$.
\begin{eqnarray*}
I_A&=\frac{\pi\Omega_{1}\Omega_{2}}{2ir}\frac{V}{(2\pi)^3}2i\pi\Delta\left(\frac{2m}{\hbar^2}\right)^2
&\left[\frac{k_1e^{ik_1r}}{(k_1-k_2)(k_1-k_3)(k_1-k_4)}+\frac{k_2e^{ik_2r}}{(k_2-k_1)(k_2-k_3)(k_2-k_4)}\right. \\
&&\left.-\frac{-k_3e^{-ik_3r}}{(k_3-k_1)(k_3-k_2)(k_3-k_4)}-\frac{-k_4e^{-ik_4r}}{(k_4-k_1)(k_4-k_2)(k_4-k_3)}\right]
\end{eqnarray*}
\begin{equation}
I_A=\frac{\pi\Omega_{1}\Omega_{2}}{2ir}\frac{V}{(2\pi)^3}\Delta\left(\frac{2m}{\hbar^2}\right)^2\frac{\pi e^{-ak_Fr\sin\theta}}{2(ak_F)^2\sin(2\theta)}\left[e^{iak_Fr\cos\theta}-e^{-iak_Fr\cos\theta}\right]
\end{equation}
with $\sin2\theta=2m\Delta/(a^2\hbar^2k_F^2)$,
$\sin\theta={\Delta}/{2E_F}$.
Given that $a$ and $\cos\theta$ $\sim1$ ($\Delta_S\ll E_F$), one obtains:
\begin{equation} \label{int_andreev}
I_A=\frac{\pi}{2}N(0)\Omega_{1}\Omega_{2}e^{-\frac{r}{\pi\xi_0}}\frac{\sin(k_Fr)}{k_Fr}
\end{equation}
In what follows, $I_A$ is noted: $I_A=\gamma_A T$ with
$T=\frac{\pi}{2}N(0)\Omega_{1}\Omega_{2}$ and
$\gamma_A=e^{-\frac{r}{\pi\xi_0}}\frac{\sin(k_Fr)}{k_Fr}$, the geometrical
factor for Crossed-Andreev effect.

\subsection{Direct-Andreev effect}

The tunneling of the two electron of a same Cooper pair to the same dot, $i$, gives a contribution
\begin{equation}
T_i=\sum_k\frac{u_kv_k\Omega_{ki}\Omega_{-ki}}{E-E_i-E_k}
\end{equation}
From the previous calculation, one must take the limit $r\rightarrow 0$ in the Eq.(\ref{int_andreev}).
The same result is found when making the calculation without taking into account the phase factor $e^{ikx}$
which generates the geometrical factor. The amplitude of this effect towards $i$-side is then 
$T_i=\frac{\pi}{2}N(0)\Omega_{i}^2$.

\subsection{Self-energy}

The self-energy terms are due to the coupling between a discrete state (state with zero or one electron in a dot) and
a continuum of states (quasiparticle states in $S$). They correspond to the renormalization of of these
energy levels. They involve $|v_k|^2$ when the annihilation of an electron in $S$ corresponds
to the creation of quasiparticle, and $|u_k|^2$ when the creation of an electron corresponds
to the creation of quasiparticle. In  Eqs. (\ref{1a}-\ref{1g}), this terms corresponds to
\begin{eqnarray}
I_R=\sum_k\frac{|v_k|^2|\Omega_{ki}|^2}{E-E_i-E_k}\\
J_R=\sum_k\frac{|u_k|^2|\Omega_{ki}|^2}{E-E_i-E_k}
\end{eqnarray}
for a given $i$-side.

The sum are transformed into integrals over quasiparticle energies, $E_k$, with a density of states given by
$N(E)=N(0)E/\sqrt{E^2-\Delta_S^2}$. For the calculation, $E_i/\Delta_S$ is noted $e_i$ and $E/\Delta_S$ is noted $x$.

\paragraph{Terms in $|v_k|^2$:}
\begin{eqnarray}
I_R&=&\frac{T_i}{2}\int_{1}^{\infty}\left(\frac{1}{\sqrt{x^2-1}}-\frac{1}{x}\right)\frac{x}{e_i-x}dx\nonumber\\
&=&-\frac{T_i}{2}\ln\left(2(1-e_i)\right)-T_i\frac{e_i}{\sqrt{e_i^2-1}}\left(\frac{\pi}{2}+\arcsin(e_i)\right)\nonumber\\
&\approx&-\frac{T_i}{2}\ln2 \nonumber\\
I_R&=&cT_i
\end{eqnarray}
where $c$ is a numerical constant.

\paragraph{Terms in $|u_k|^2$:} This term never appears alone, so we just
have to calculate terms with $|u_k|^2-|v_k|^2$.
\begin{eqnarray}
J_R-I_R&=&\frac{T_i}{2}\int_{1}^{\infty}\frac{dx}{e_i-x}\nonumber \\
&=&-\frac{T_i}{2}\left[\ln(e_i-x)\right]_1^{\infty}\nonumber \\
J_R&=&c'T_i
\end{eqnarray}
To avoid the logarithmic divergence we introduce a physical cutoff -- the electron band width -- to get a finite result.
This does not yield a large contribution because of the logarithm: if the band width is 1000 times higher than the gap 
it only gives a factor $8\Omega^2$ where $\Omega\ll E_i$. Self energy terms remain small. Let us define $K=c'-2c'$ for 
the following.

\subsection{S--Cotunneling}

Local S--cotunneling has no meaning (tunneling of an electron between two places) so keeping
the geometrical contribution of the integrand in this process, one gets:
\begin{equation}
I_C=\sum_k\frac{(|u_k|^2-|v_k|^2)\Omega_{k1}\Omega_{-k2}}{E-E_i-E_k}
\end{equation}
With $|u_k|^2-|v_k|^2=\xi_k/E_k$:
\begin{eqnarray}
I_C&=&{\displaystyle
\frac{\pi\Omega_{1}\Omega_{2}}{2ir}\frac{V}{(2\pi)^3}\int_{-\infty}^{\infty}dk\,k\left(e^{ikr}-e^{-ikr}\right)\frac{\frac{\hbar^2k^2}{2m}-\mu}{\Delta^2+(\frac{\hbar^2k^2}{2m}-\mu)^2}} \nonumber
\end{eqnarray}
Using once again the residue theorem one gets:
\begin{equation}
I_C=\frac{\pi\Omega_{1}\Omega_{2}}{2ir}\frac{V}{(2\pi)^3}\left(\frac{2m}{\hbar^2}\right)^2\frac{\pi k_F^2e^{-ak_Fr\sin\theta}}{(ak_F)^2\sin(2\theta)}\left[e^{i(ak_Fr\cos\theta+\theta)}\left(a^2e^{i\theta}-e^{-i\theta}\right)-e^{-i(ak_Fr\cos\theta+\theta)}\left(a^2e^{-i\theta}-e^{i\theta}\right)\right]\end{equation}
With $a^2\sim1+\frac{1}{2}\left(\frac{2m\Delta}{\hbar^2k_F^2}\right)^2$:

\begin{equation}
I_C=\frac{\pi}{2}N(0)\Omega_{1}\Omega_{2}e^{-\frac{r}{\pi\xi_0}}\left(\frac{\cos(k_Fr)}{k_Fr}+\frac{1}{2}\left(\frac{2m\Delta}{\hbar^2k_F^2}\right)^2\left(\frac{\sin(k_Fr)}{k_Fr}-\frac{\cos(k_Fr)}{k_Fr}\right)\right)
\end{equation}

The second term is much smaller than the first one ($\Delta_S\ll E_F$). The only difference with the Andreev amplitude is the
$\frac{\cos(k_Fr)}{k_Fr}$ instead of $\frac{\sin(k_Fr)}{k_Fr}$. S--cotunneling diverges for $r\rightarrow0$.

\subsection{One-by-one electron tunneling to the reservoir}
\label{Abranche}

Here the calculation is not complicated by a phase factor.
The sum over $k$ is simply replaced by an integral over energy.
\begin{eqnarray}
I_P&=&{\displaystyle \sum_k \frac{u_kv_k|\Omega_{k1}|^2}{\Delta_{lk}(\Delta_{1k+i\Gamma_L/2})}}\nonumber \\
&\simeq&{\displaystyle \sum_k\frac{u_kv_k}{E_k^2}\Omega_1^2\Omega_l}\nonumber \\
&=&{\displaystyle
N(0)\Omega_1^2\Omega_l\int_{\Delta}^{\infty}\frac{E}{\sqrt{E^2-\Delta^2}}\frac{\Delta}{E^3}dE}\nonumber \\
&=&{\displaystyle N(0)\frac{\Omega_1^2\Omega_l}{\Delta}\int_{1}^{\infty}\frac{dx}{x^2\sqrt{x^2-1}}}\nonumber \\
&=&{\displaystyle N(0)\frac{\Omega_1^2\Omega_l}{\Delta}} \label{one-by-one}
\end{eqnarray}

\section{Quantum master equations for the Entangler} \label{system}

The set of fully consistent and non-perturbative quantum master equations can be derived (see main part of the paper).
For simplicity, the space of charge states has been restricted here to 0, 1, 2 or 3 electrons in the two dots. 
Numerical calculations have been made with this set of equations including states $|g\rangle$
(one electron in dot 2, two electrons in dot 1), $|h\rangle$ (one electron in dot 1, two electrons in dot 2) and 
$|k\rangle$ (triplet state shared between dots 1 and 2). 

\begin{equation}
\begin{array}{rl}
\dot{\sigma}_{aa}=&{\displaystyle +2iT_1\left(\sigma_{ae}-\sigma_{ea}\right)+2iT_2\left(\sigma_{af}-\sigma_{fa}\right)}\\
&{\displaystyle 2\sqrt{2}i\gamma_A T\left(\sigma_{ab}-\sigma_{ba}\right)+\Gamma_L\sigma_{cc}+\Gamma_R\sigma_{dd}-2\left(\tilde{\Gamma}_L+\tilde{\Gamma}_R\right)\sigma_{aa}}
\end{array}
\end{equation}
\begin{equation}
\begin{array}{rl}
\dot{\sigma}_{bb}=&+i\sqrt{2}\gamma_CT\left(\sigma_{be}-\sigma_{eb}\right)+i\sqrt{2}\gamma_CT\left(\sigma_{bf}-\sigma_{fb}\right)-2\sqrt{2}i\gamma_AT\left(\sigma_{ab}-\sigma_{ba}\right) \\
&+\frac{1}{2}\tilde{\Gamma}_R\sigma_{cc}+\frac{1}{2}\tilde{\Gamma}_L\sigma_{dd}+\frac{1}{2}\Gamma'_L\sigma_{gg}+\frac{1}{2}\Gamma'_R\sigma_{hh}-2\left(\Gamma_L+\Gamma_R\right)\sigma_{bb}
\end{array}
\end{equation}
\begin{equation}
\dot{\sigma}_{cc}=i\gamma_CT\left(\sigma_{cd}-\sigma_{dc}\right)+2iT_2\left(\sigma_{ch}-\sigma_{hc}\right)+2\tilde{\Gamma}_L\sigma_{aa}+2\Gamma_R\sigma_{bb}+2\Gamma'_L\sigma_{ee}-(\Gamma_L+2\tilde{\Gamma}_R)\sigma_{cc}
\end{equation}
\begin{equation}
\dot{\sigma}_{dd}=i\gamma_CT\left(\sigma_{cd}-\sigma_{dc}\right)+2iT_1\left(\sigma_{dg}-\sigma_{gd}\right)+2\tilde{\Gamma}_R\sigma_{aa}+2\Gamma_L\sigma_{bb}+2\Gamma'_R\sigma_{ff}-(\Gamma_R+2\tilde{\Gamma}_L)\sigma_{dd}
\end{equation}
\begin{equation}
\dot{\sigma}_{ee}=-2iT_1\left(\sigma_{ae}-\sigma_{ea}\right)-i\sqrt{2}\gamma_CT\left(\sigma_{be}-\sigma_{eb}\right)+\Gamma_R\sigma_{gg}-2\Gamma'_L\sigma_{ee}
\end{equation}
\begin{equation}
\dot{\sigma}_{ff}=-2iT_2\left(\sigma_{af}-\sigma_{fa}\right)-i\sqrt{2}\gamma_CT\left(\sigma_{bf}-\sigma_{fb}\right)+\Gamma_L\sigma_{hh}-2\Gamma'_R\sigma_{ff}
\end{equation}
\begin{equation}
\dot{\sigma}_{gg}=-2iT_1\left(\sigma_{dg}-\sigma_{gd}\right)-(2\Gamma'_L+\Gamma_R)\sigma_{gg}
\end{equation}
\begin{equation}
\dot{\sigma}_{hh}=-2iT_2\left(\sigma_{ch}-\sigma_{hc}\right)-(2\Gamma'_R+\Gamma_L)\sigma_{hh}
\end{equation}
\begin{equation}
\dot{\sigma}_{kk}=\frac{3}{2}(\tilde{\Gamma}_R\sigma_{cc}+\tilde{\Gamma}_L\sigma_{dd}+\Gamma'_L\sigma_{gg}+\Gamma'_R\sigma_{hh})-(\Gamma_L+\Gamma_R)\sigma_{hh}
\end{equation}
\begin{equation}
\begin{array}{rl}
\dot{\sigma}_{ab}=&i\left(E'_1+E'_2\right)\sigma_{ab}+2\sqrt{2}i
\gamma_AT\left(\sigma_{aa}-\sigma_{bb}\right)+i\sqrt{2}T(\sigma_{ae}+\sigma_{af})-(iT_1\sigma_{eb}+iT_2\sigma_{fb})\\
&-\frac{1}{2}\left(2\tilde{\Gamma}_L+2\tilde{\Gamma}_R+\Gamma_L+\Gamma_R\right)\sigma_{ab}
\end{array}
\end{equation}
\begin{equation}
\dot{\sigma}_{ae}=i(E'_1+U'_{11})\sigma_{ae}+2iT_1\left(\sigma_{aa}-\sigma_{ee}\right)+i(\sqrt{2}\gamma_CT\sigma_{ab}-2\sqrt{2}\gamma_AT\sigma_{be}-2T_2\sigma_{fe})-(\tilde{\Gamma}_L+\tilde{\Gamma}_R+\Gamma'_L)\sigma_{ae}
\end{equation}
\begin{equation}
\dot{\sigma}_{af}=i(E'_2+U'_{22})\sigma_{af}+2iT_2\left(\sigma_{aa}-\sigma_{ff}\right)+i(\sqrt{2}\gamma_CT\sigma_{ab}-2\sqrt{2}\gamma_AT\sigma_{bf}-2T_1\sigma_{ef})-(\tilde{\Gamma}_L+\tilde{\Gamma}_R+\Gamma'_R)\sigma_{af}
\end{equation}
\begin{equation}
\dot{\sigma}_{be}=i\left(U'_{11}-E'_2\right)\sigma_{be}+i\sqrt{2}\gamma_CT\left(\sigma_{bb}-\sigma_{ee}\right)+i(2T_1\sigma_{ba}-2\sqrt{2}\gamma_AT\sigma_{ae}-\sqrt{2}\gamma_CT\sigma_{fe})-\left(\Gamma_L+\Gamma_R+\Gamma'_L\right)\sigma_{be}
\end{equation}
\begin{equation}
\dot{\sigma}_{bf}=i\left(U'_{22}-E'_1\right)\sigma_{bf}+i\sqrt{2}\gamma_CT\left(\sigma_{bb}-\sigma_{ff}\right)+i(2T_2\sigma_{ba}-2\sqrt{2}\gamma_AT\sigma_{af}-\sqrt{2}\gamma_CT\sigma_{ef})-\left(\Gamma_L+\Gamma_R+\Gamma'_R\right)\sigma_{bf}
\end{equation}
\begin{equation}
\dot{\sigma}_{cd}=i\left(E'_2-E'_1\right)\sigma_{cd}+i\gamma_CT\left(\sigma_{cc}-\sigma_{dd}\right)+2i(T_1\sigma_{cg}-T_2\sigma_{hd})-\frac{1}{2}\left(\Gamma_L+\Gamma_R+2\tilde{\Gamma}_L+2\tilde{\Gamma}_R\right)\sigma_{cd}
\end{equation}
\begin{equation}
\dot{\sigma}_{ch}=i\left(U'_{22}+E'_2\right)\sigma_{ch}+2iT_2(\sigma_{cc}-\sigma_{hh})-i\gamma_CT\sigma_{dh}-(\Gamma_L+\Gamma'_R+\tilde{\Gamma}_R)\sigma_{ch}
\end{equation}
\begin{equation}
\dot{\sigma}_{dg}=i\left(U'_{11}+E'_1\right)\sigma_{dg}+2iT_1(\sigma_{dd}-\sigma_{gg})-i\gamma_CT\sigma_{cg}-(\Gamma_R+\Gamma'_L+\tilde{\Gamma}_L)\sigma_{ch}
\end{equation}
where $E'$ and $U'$ are the renormalized energy levels. $\tilde{\Gamma}_{L,R}=K''^2\Gamma_{L,R} T_{1,2}/\Delta_S$ corresponds
to the one-by-one process. The superconducting phase, which do not change any result, is omitted here.

\section{Derivation of quantum master equations for the Entangler using a pair-Hamiltonian}

The same set of quantum master equations could be obtained from an effective Hamiltonian applied to the
method developed in Ref.\onlinecite{gurvitz_prager}. In considering all processes, this Hamiltonian can be 
derived from the microscopic Hamiltonian (eq.\ref{hamiltonian}) using a projective transformation which 
eliminates states with quasiparticles in the superconductor to the lowest order:
\begin{eqnarray}
\label{effective}
\mathcal{H}_{eff}=P\mathcal{H}_0P+2\sqrt{2}\gamma_AT\frac{1}{\sqrt{2}}\left(d_{1\sigma}^{\dagger}d_{2-\!\sigma}^{\dagger}-d_{1-\!\sigma}^{\dagger}d_{2\sigma}^{\dagger}\right)S
\nonumber\\
+2T_1(d_{1\sigma}^{\dagger}d_{1-\!\sigma}^{\dagger})S+2T_2(d_{2\sigma}^{\dagger}d_{2-\!\sigma}^{\dagger})S+\sqrt{2}\gamma_CT\sum_{\sigma}d_{1\sigma}^{\dagger}d_{2\sigma}\nonumber\\
+\sum_{l,\sigma}\Omega_{l}a_{l\sigma}^{\dagger}d_{1\sigma}+\sum_{r,\sigma}\Omega_{r}a_{r\sigma}^{\dagger}d_{2\sigma}+\sum_{l,\sigma}\widehat{\Omega}_{l}d_{1\sigma}^{\dagger}a_{l\sigma}^{\dagger}S+\sum_{r,\sigma}\widehat{\Omega}_{r}d_{2\sigma}^{\dagger}a_{r\sigma}^{\dagger}S+h.c.
\end{eqnarray}
with $\widehat{\Omega}_{l,r}=\Omega_{l,r}T_{1,2}/\Delta_S$ corresponds to the one-by-one process. The method only 
requires the amplitude for probability of processes coupling different states of the quantum system, and leads to 
a following general system:
\begin{equation}
\dot{\sigma}_{\alpha\alpha}=i\sum_{\gamma\neq\alpha}\Omega_{\alpha\gamma}\left(\sigma_{\alpha\gamma}-\sigma_{\gamma\alpha}\right)-\sum_{\gamma\neq\alpha}\Gamma_{\alpha\rightarrow\gamma}\sigma_{\alpha\alpha}+\sum_{\gamma\neq\alpha}\Gamma_{\gamma\rightarrow\alpha}\sigma_{\gamma\gamma}
\end{equation}
\begin{eqnarray}
\dot{\sigma}_{\alpha\beta}=i(E_{\beta}-E_{\alpha})\sigma_{\alpha\beta}+i\sum_{\gamma\neq\beta}\left(\sigma_{\alpha\gamma}\Omega_{\gamma\beta}-i\sum_{\gamma\neq\alpha}\Omega_{\alpha\gamma}\sigma_{\gamma\beta}\right)\nonumber\\
-\frac{1}{2}\left(\sum_{\gamma\neq\alpha}\Gamma_{\alpha\rightarrow\gamma}-\sum_{\gamma\neq\beta}\Gamma_{\beta\rightarrow\gamma}\right)\sigma_{\alpha\beta}+\frac{1}{2}\sum_{\gamma\delta\neq\alpha\beta}\left(\Gamma_{\gamma\rightarrow\alpha}+\Gamma_{\delta\rightarrow\beta}\right)\sigma_{\gamma\delta}
\end{eqnarray}
where the $\Omega$'s are the coherent transition matrix elements and the $\Gamma$'s the relaxation rates.

\end{document}